\documentclass[aps,prc,twocolumn,showpacs,preprintnumbers,
               nofootinbib,float,superscriptaddress,longbibliography]{revtex4-2}
\usepackage{graphicx}
\usepackage{color}
\usepackage[dvipsnames]{xcolor}
\usepackage[colorlinks=true, pdfstartview=FitV, linkcolor=RedOrange, citecolor=ForestGreen, urlcolor=blue]{hyperref}
\usepackage{amsmath}
\usepackage{amssymb}
\usepackage{bm}
\usepackage{physics}
\usepackage{placeins}
\usepackage{adjustbox}

\newcommand{\bs}{\boldsymbol}

\usepackage[maxfloats=256]{morefloats}
\begin{document}
\title{Bayesian analysis of (3+1)D relativistic nuclear dynamics with the RHIC beam energy scan data}

\author{Syed Afrid Jahan} \email{hm0746@wayne.edu}
\affiliation{Department of Physics and Astronomy, Wayne State University, Detroit, Michigan 48201, USA}
\author{Hendrik Roch} \email{Hendrik.Roch@wayne.edu}
\affiliation{Department of Physics and Astronomy, Wayne State University, Detroit, Michigan 48201, USA}
\author{Chun Shen} \email{chunshen@wayne.edu}
\affiliation{Department of Physics and Astronomy, Wayne State University, Detroit, Michigan 48201, USA}

\begin{abstract}
This work presents a Bayesian inference study for relativistic heavy-ion collisions in the Beam Energy Scan program at the Relativistic Heavy-Ion Collider.
The theoretical model simulates event-by-event (3+1)D collision dynamics using hydrodynamics and hadronic transport theory.
We analyze the model's 20-dimensional posterior distributions obtained using three model emulators with different accuracy and demonstrate the essential role of training an accurate model emulator in the Bayesian analysis.
Our analysis provides robust constraints on the Quark-Gluon Plasma's transport properties and various aspects of (3+1)D relativistic nuclear dynamics.
By running full model simulations with 100 parameter sets sampled from the posterior distribution, we make predictions for $p_{\rm T}$-differential observables and estimate their systematic theory uncertainty.
A sensitivity analysis is performed to elucidate how individual experimental observables respond to different model parameters, providing useful physics insights into the phenomenological model for heavy-ion collisions. 
\end{abstract}
\maketitle

\section{Introduction}
\label{sec:intro}
Relativistic heavy-ion collision programs at the Relativistic Heavy Ion Collider (RHIC) and the Large Hadron Collider (LHC) have been a key tool in studying the properties of the Quark Gluon Plasma (QGP), a state of matter exhibiting quarks and gluon degrees of freedom~\cite{Achenbach:2023pba,Arslandok:2023utm}.
These experiments allow high-energy nuclear physicists to quantitatively characterize the strongly interacting nuclear matter under extreme temperatures and densities, thereby exploring the phase structure of Quantum Chromodynamics (QCD).
Especially with the systems studied in the RHIC Beam Energy Scan (BES) program, where the center-of-mass energy is varied, it is possible to study the QGP over a wide range of temperatures and densities~\cite{Caines:2009yu,Mohanty:2011nm,Mitchell:2012mx,Odyniec:2015iaa}.
With precision measurements from this experimental program, it is exciting to investigate the transition between the QGP and hadronic matter, search for a critical point and the associated first-order phase boundary in the QCD phase diagram, and study the emergent properties of the produced state of matter (see reviews~\cite{STAR:2010vob,Luo:2017faz,Bzdak:2019pkr,Shen:2020mgh,An:2021wof}).

Modeling heavy-ion collisions event-by-event is a complex and computationally demanding task~\cite{Shen:2014vra,Putschke:2019yrg,Schenke:2020mbo,Nijs:2020roc,Pang:2018zzo, Shen:2022oyg}.
Because the relevant length scales change dynamically as the collision systems evolve, it requires matching different types of physical models in a multi-stage approach.
Fluid dynamics serves as an effective long-wavelength description of the collective behavior of the strongly coupled QGP produced in heavy-ion collisions.
Relativistic viscous hydrodynamics can well describe the time evolution of the produced bulk medium, efficiently transforming the initial-state spatial inhomogeneities to anisotropies in the final-state hadrons momentum distributions. This theoretical description allows us to access information about the initial conditions of heavy-ion collisions and transport properties of the QGP from the experimental measurements.

It is challenging to compute the QGP transport coefficients from first principles~\cite{Moore:2020pfu}. There have been several attempts using lattice QCD techniques to compute the plasma's shear viscosity for a pure gluon system~\cite{Nakamura:2004sy, Meyer:2007ic, Astrakhantsev:2017nrs, Altenkort:2022yhb}.
On the other hand, extensive phenomenological studies were able to show that the hadronic observables measured from heavy-ion collisions are sensitive to the viscosities in the QGP~\cite{Schenke:2020mbo,Song:2010mg,Karpenko:2015xea,Shen:2015msa,Ryu:2015vwa,Schenke:2019ruo,Ryu:2017qzn}.
There have been many studies using hydrodynamics in heavy-ion collision simulations to estimate the specific shear viscosity $\eta/s$ for the QGP~\cite{Song:2010mg,Karpenko:2015xea,Niemi:2015qia,Shen:2020jwv, Shen:2020gef}.
Nowadays, there is more effort put into the direction of extracting the QGP-specific shear and bulk viscosities, together with their uncertainties from varying other aspects of the theoretical model, based on large-scale model-to-data comparisons with Bayesian inference analysis in a high dimensional model parameter space~\cite{Pratt:2015zsa,Bernhard:2016tnd,Auvinen:2017fjw,Bernhard:2019bmu,Nijs:2020ors,JETSCAPE:2020mzn,JETSCAPE:2020shq,Parkkila:2021tqq,Parkkila:2021yha,Phillips:2020dmw,Heffernan:2023gye,Heffernan:2023utr, JETSCAPE:2023nuf}.

The RHIC BES program generates a wealth of data for heavy-ion collisions at different collision energies. These measurements enable phenomenological extraction of the QGP transport properties at finite baryon density via the global Bayesian inference analysis.
For collisions with energies around $\mathcal{O}(10)$ GeV, full (3+1)D simulations are required to perform precision comparisons~\cite{Shen:2017fnn, Shen:2021nbe}. 

This work expands upon Ref.~\cite{Shen:2023awv} and performs a systematic Bayesian inference analysis of RHIC BES measurements using full (3+1)D event-by-event simulations. In addition to the constraints on the QGP-specific viscosity, we will report and analyze the posterior distribution for all model parameters in Sec.~\ref{sec:posterior}.
We improve the constraints on the model parameters by applying a more accurate Gaussian Process (GP) model emulator~\cite{Roch:2024xhh} than the one used in~\cite{Shen:2023awv}. Sec.~\ref{sec:model_sim} will further provide a selection of model predictions by performing numerical simulations with the parameter sets drawn from the posterior distribution. The relative variations in the model predictions from different parameter sets provide the theoretical uncertainty from the Bayesian calibration. We will perform a detailed sensitivity analysis between the model parameters and experimental observables in Sec.~\ref{sec:sensitivity}.

\section{The Hybrid Theoretical Framework and Model Parameters}
\label{sec:framework_parameters}

The theoretical simulations for Au+Au collisions at $\sqrt{s_{\rm NN}}=7.7, 19.6$, and $200$~GeV used for this Bayesian analysis were performed using the \texttt{iEBE-MUSIC} framework~\cite{Shen:2023awv}.
To simulate the full (3+1)D dynamical evolution, a \texttt{3D-Glauber} model is coupled with relativistic viscous hydrodynamics (\texttt{MUSIC})~\cite{Schenke:2010nt, Paquet:2015lta, Denicol:2018wdp} and hadronic transport (\texttt{UrQMD})~\cite{Bass:1998ca, Bleicher:1999xi}.
The \texttt{3D-Glauber} model considers the finite thickness of nuclei along the beam direction as they pass through each other, which becomes important as the center of mass collision energy decreases to $\mathcal{O}(10)$~GeV~\cite{Shen:2017ruz, Shen:2017bsr, Shen:2023aeg}. The transverse collision geometry is controlled by the shape of nuclei and their relative impact parameter. Each nucleon inside the colliding nuclei contains four hotspots related to three constituent quarks and one soft gluon. These hotspots lose energy and longitudinal momentum by being decelerated by strings between them during nucleon-nucleon collisions~\cite{Shen:2022oyg}.
The averaged rapidity loss is parametrized as a function of the incoming rapidity $y_{\rm init}$ in the collision pair rest frame~\cite{Shen:2023awv}:
\begin{equation}
    \langle y_{\rm loss}\rangle = \begin{cases}
        y_{\rm loss,2}\frac{y_{\rm init}}{2},\; 0<y_{\rm init}\leq 2\\
        y_{\rm loss,2} + (y_{\rm loss,4} - y_{\rm loss,2})\frac{y_{\rm init}-2}{2},\; 2<y_{\rm init}<4\\
        y_{\rm loss,4} + (y_{\rm loss,6} - y_{\rm loss,4})\frac{y_{\rm init}-4}{2},\; y_{\rm init}\geq 4
    \end{cases}
    \label{eq:ylossParam}
\end{equation}
where $y_{{\rm loss}, n}$ is the average amount of rapidity loss for $y_{\rm init}=n$.
The event-by-event fluctuations of the rapidity loss are modeled by the parameter $\sigma_{y_{\rm loss}}$, which controls the variance of the fluctuations~\cite{Shen:2022oyg}.
The parameter $\alpha_{\rm shadowing}$ takes into account coherence effects when the incoming nucleons go through multiple collisions.
With $\alpha_{\rm shadowing}=0$, all binary collisions will produce strings while $\alpha_{\rm shadowing} \neq 0$ sets a finite probability to shadow string production from secondary collisions between nucleon pairs.
The parameter $\alpha_\mathrm{rem}$ controls the amount of energy-momentum lost by the collision remnants~\cite{Shen:2022oyg}.
These wounded partons are connected via strings and decelerated in the longitudinal direction until $\tau_{\rm hydro}=0.5\;\mathrm{fm}/c$ in the collision rest frame.
Then, they are fed as source terms for the system's energy-momentum tensor and net baryon current,
\begin{align}
    \partial_\mu T^{\mu\nu} &= J^\nu, \label{eq:hydroEOM1} \\
    \partial_\mu J^\mu_B &= \rho_B.  \label{eq:hydroEOM2} 
\end{align}
The definitions of the source terms were explained in detail in Ref.~\cite{Shen:2022oyg}.
Here, we introduce a longitudinal tilted string profile by parameterizing the transverse profile as,
\begin{align}
    & f(\vec{x}_\perp, \eta_s) = \frac{1}{2\pi (\sigma_x^\mathrm{string})^2 }  \nonumber \\ 
    & \qquad \qquad \exp \left[ - \frac{[x - x_c(\eta_s)]^2 + [y - y_c(\eta_s)]^2}{2 (\sigma_x^\mathrm{string})^2} \right],
\end{align}
where the position of the string center $(x_c, y_c)$ depends on the space-time rapidity,
\begin{align}
    x_c (\eta_s) &= \frac{x^P + x^T}{2} + \alpha_{\text{string tilt}} \frac{x^P - x^T}{\eta_s^P - \eta_s^T}\left(\eta_s - \frac{\eta_s^P + \eta_s^T}{2}\right), \label{eq:stringCenter_x} \\
    y_c (\eta_s) &= \frac{y^P + y^T}{2} + \alpha_{\text{string tilt}} \frac{y^P - y^T}{\eta_s^P - \eta_s^T}\left(\eta_s - \frac{\eta_s^P + \eta_s^T}{2}\right). \label{eq:stringCenter_y}
\end{align}
Here, $(x^{P/T}, y^{P/T}, \eta_s^{P/T})$ are the space-time coordinates for the projectile ($P$) and target ($T$) hotspots after the string deceleration.

Motivated by the baryon junction model~\cite{Kharzeev:1996sq}, we allow for the incoming baryon numbers to be located at the string ends or fluctuate along the string with a parameter $\lambda_B$~\cite{Shen:2022oyg},
\begin{equation}
    P(y^B_{P/T}) = (1 - \lambda_B) y_{P/T} + \lambda_B \frac{e^{y^B_{P/T} - (y_P + y_T)/2}}{4 \sinh[(y_P - y_T)/4]}.
\end{equation}

The pre-equilibrium evolution of the system is modeled via a blast-wave-like transverse flow profile for the individual strings developed in the time $\leq\tau_{\rm hydro}$ with a transverse flow rapidity~\cite{Zhao:2022ugy, Shen:2023awv}
\begin{equation}
    \eta_\perp(\mathbf{x}_\perp) = \alpha_{\rm preFlow}|\tilde{\mathbf{x}}_\perp|,
\end{equation}
where $\tilde{\mathbf{x}}_\perp=(x-x_c, y-y_c)$, with $x_c$ and $y_c$ being the string's center in the transverse plane in Eqs.~\eqref{eq:stringCenter_x} and \eqref{eq:stringCenter_y}.

The hydrodynamic equation of motion, Eqs.~\eqref{eq:hydroEOM1} and \eqref{eq:hydroEOM2}, are evolved with the lattice-QCD-based \texttt{NEOS-BQS} equation of state at finite densities, which assumes strangeness neutrality and $n_Q=0.4n_B$ for Au+Au collisions~\cite{Monnai:2019hkn}.
The system's viscous stress tensors are evolved with the Denicol-Niemi-Molnar-Rischke (DNMR) theory~\cite{Denicol:2012cn}.
The $\mu_B$ dependence of QGP-specific shear viscosity is parametrized by 
\begin{equation}
    \tilde{\eta}(\mu_B)=\begin{cases}
        \eta_0 + (\eta_2 - \eta_0)\frac{\mu_B}{0.2},\; 0<\mu_B\leq 0.2\;\mathrm{GeV}\\
        \eta_2 + (\eta_4 - \eta_2)\frac{(\mu_B - 0.2)}{0.2},\; 0.2<\mu_B<0.4\;\mathrm{GeV},\\
        \eta_4,\; \mu_B\geq 0.4\;\mathrm{GeV}
    \end{cases}
    \label{eq:QGPshearParam}
\end{equation}
where $\tilde{\eta}\equiv\eta T/(e+P)$ and $\eta_0$, $\eta_2$ and $\eta_4$ are the values of the specific shear viscosity at $\mu_B=0,0.2,0.4\;\mathrm{GeV}$ respectively. We did not introduce an explicit temperature dependence in the parameterization to keep the number of model parameters manageable. A previous analysis showed that a temperature-independent $\eta/s$ is compatible with experimental data at high energy collisions~\cite{JETSCAPE:2020mzn}.
We parameterize the specific bulk viscosity by
\begin{equation}
    \tilde{\zeta}(T,\mu_B)=\begin{cases}
        \zeta_{\rm max}\exp\left[-\frac{[T - T_\zeta(\mu_B)]^2}{2\sigma^2_{\zeta,-}}\right],\; T<T_\zeta(\mu_B) \\
        \zeta_{\rm max}\exp\left[-\frac{[T - T_\zeta(\mu_B)]^2}{2\sigma^2_{\zeta,+}}\right],\; T\geq T_\zeta(\mu_B)
    \end{cases},
    \label{eq:QGPbulkParam}
\end{equation}
where $\tilde{\zeta}\equiv \zeta T/(e+P)$ with the bulk pressure peak temperature $T_\zeta(\mu_B)\equiv T_{\zeta,0}-\frac{0.15}{1\;\mathrm{GeV}}\mu_B^2$.
This parameterization follows the constant energy density curve $e=e(T_{\zeta,0},\mu_B=0)$ of the \texttt{NEOS-BQS} equation of state closely and ensures that the parameterization peaks close to the phase crossover at $\mu_B>0$~\cite{An:2021wof,HotQCD:2018pds,Borsanyi:2020fev}. 

We choose to normalize the QGP shear and bulk viscosity by $T/(e + P)$ instead of the local entropy density $s$ to better control the variations of the viscous relaxation times in numerical simulations. Take the shear relaxation time, for example,
\begin{align}
    \tau_\pi \equiv C_\eta \frac{\eta}{(e + P)} &= C_\eta \tilde{\eta} \frac{1}{T}  \label{eq:taupi1} \\
    &= C_\eta \frac{\eta}{s} \frac{1}{T} \frac{1}{1 + \frac{\mu_B n_B}{Ts}}.\label{eq:taupi2}
\end{align}
One can see that the shear relaxation time $\tau_\pi$ in Eq.~\eqref{eq:taupi1} remains well behaved when we vary $\tilde{\eta}$ within the prior range. If we parameterize $\eta/s$ as in Eq.~\eqref{eq:QGPshearParam}, then $\tau_\pi$ in Eq.~\eqref{eq:taupi2} could potentially reduce to very small values and violate causality conditions~\cite{Hiscock:1983zz, Chiu:2021muk} when the factor $\mu_B n_B/(Ts)$ becomes large during the dynamical evolution.

All fluid cells dropping below an energy density $e_{\rm sw}$ are particlized using a Cooper-Frye freeze-out procedure including $\delta f$ corrections for multiple conserved charge currents ($B, Q, S$) using the Grad moment method~\cite{Huovinen:2012is, Shen:2014vra}.
The hadronic rescatterings and decays in the afterburner phase are modeled using \texttt{UrQMD}~\cite{Bass:1998ca,Bleicher:1999xi}.

All 20 model parameters, together with their prior ranges, are listed in Table~\ref{tab:parameters}.

\begin{table}[h!]
    \caption{The 20 model parameters and their prior ranges.}
    \label{tab:parameters}
    \begin{tabular}{c|c|c|c}
    \hline\hline
    Parameter & Prior & Parameter & Prior \\
    \hline
    $B_G\;[\mathrm{GeV}^{-2}]$ & $[1,25]$ & $\alpha_{\text{string tilt}}$ & $[0,1]$ \\
    $\alpha_{\rm shadowing}$ & $[0,1]$ & $\alpha_{\text{preFlow}}$ & $[0,2]$ \\
    $y_{{\rm loss},2}$ & $[0,2]$ & $\eta_0$ & $[0.001,0.3]$ \\
    $y_{{\rm loss},4}$ & $[1,3]$ & $\eta_2$ & $[0.001,0.3]$ \\
    $y_{{\rm loss},6}$ & $[1,4]$ & $\eta_4$ & $[0.001,0.3]$ \\
    $\sigma_{y_{\rm loss}}$ & $[0.1,0.8]$ & $\zeta_{\rm max}$ & $[0,0.2]$ \\
    $\alpha_{\rm rem}$ & $[0,1]$ & $T_{\zeta,0}\;[{\rm GeV}]$ & $[0.15,0.25]$\\
    $\lambda_B$ & $[0,1]$ & $\sigma_{\zeta,+}\;[{\rm GeV}]$ & $[0.01,0.15]$ \\
    $\sigma_x^{\rm string}\;[{\rm fm}]$ & $[0.1,0.8]$ & $\sigma_{\zeta,-}\;[{\rm GeV}]$ & $[0.005,0.1]$ \\
    $\sigma_\eta^{\rm string}$ & $[0.1,1]$ & $e_{\rm sw}\;[{\rm GeV}/{\rm fm}^3]$ & $[0.15,0.5]$\\
    \hline\hline
    \end{tabular}
\end{table}

\section{Model Emulation and Bayesian Inference Setup}
\label{sec:Bayes}
For the exploration of the 20-dimensional parameter space shown in Table~\ref{tab:parameters} with Bayesian inference, we need efficient Gaussian Process (GP) emulators to replace the computationally expensive high-fidelity model calculations with the \texttt{iEBE-MUSIC} framework.
In this work, we will make use of the PCSK emulator implemented in the \texttt{surmise} package~\cite{surmise2023} and a standard GP emulator from the Scikit-learn Python package~\cite{Shen:2023awv}.
For the details of the implementation of these emulators, we refer to our previous comparison of these GP emulators in Ref.~\cite{Roch:2024xhh}.

For an efficient exploration of the parameters in Table~\ref{tab:parameters}, we generate model simulations at 1,000 different parameter sets in 20 dimensions using the Maximum Projection Latin Hypercube Design (LHD). At every design point, the model simulations consist of 1,000 minimum bias Au+Au collisions at $200\;\mathrm{GeV}$ and 2,000 minimum bias events each at $19.6\;\mathrm{GeV}$ and $7.7\;\mathrm{GeV}$.
Additionally, we make use of 95 additional full model simulations, where the parameters were sampled from the posterior distribution.
This additional dataset is denoted as High Probability Posterior (HPP) points.
In our standard setup, we train the GP emulator with the LHD + HPP datasets combined, which improves the emulator predictions in the parameter region with higher probability.

The fast prediction of model outputs at different points in the parameter space using GP emulators allows us to make use of Bayes' rule:
\begin{equation}
\mathcal{P}(\bs{\theta} | {\bf y}_{\rm exp}) = \frac{\mathcal{P}({\bf y}_{\rm exp} | \bs{\theta}) \mathcal{P}(\bs{\theta})}{\mathcal{P}({\bf y}_{\rm exp})}.
\label{eq:bayes}
\end{equation}
$\mathcal{P}(\bs{\theta} | {\bf y}_{\rm exp})$ represents the posterior probability density function, $\mathcal{P}(\bs{\theta})$ is the prior probability density function, $\mathcal{P}({\bf y}_{\rm exp})$ stands for the total evidence, and $\mathcal{P}({\bf y}_{\rm exp} | \bs{\theta})$ is the likelihood function.
In this work, we choose uniform prior distributions for all the model parameters listed in Table~\ref{tab:parameters}.
The likelihood function is defined as,
\begin{align}
&\mathcal{P}({\bf y}_{\rm exp} | \bs{\theta}) = \frac{1}{\sqrt{|2\pi\bs{\Sigma}|}} \nonumber \\
& \quad \times \exp\left[-\frac{1}{2}({\bf y}_{\rm sim}(\bs{\theta})-{\bf y}_{\rm exp})^\mathsf{T}\bs{\Sigma}^{-1}({\bf y}_{\rm sim}(\bs{\theta})-{\bf y}_{\rm exp})\right],
\label{eq:likelihood}
\end{align}
where $\bs{\Sigma} \equiv \bs{\Sigma}_\mathrm{model} + \bs{\Sigma}_\mathrm{exp}$ is the covariance matrix. The GP emulator provides the model covariance matrix $\bs{\Sigma}_\mathrm{model}$. Lacking the knowledge of the full covariance matrix for the experimental measurements, we assume a diagonal form for $\bs{\Sigma}_\mathrm{exp}$ using the square of the statistical errors for its elements. 

To tackle the computation of the posterior distribution numerically, we employ Markov Chain Monte Carlo (MCMC) techniques~\cite{brooks2011handbook,Trotta2008}.
In Ref.~\cite{Roch:2024xhh}, we have shown that the affine invariant MCMC implemented in the \texttt{emcee} Python package~\cite{Foreman_Mackey_2013} and the Parallel Tempering Langevin Monte Carlo (PTLMC) implemented in \texttt{surmise} lead to consistent results.
The parallel tempering methods generally work better when there are multiple modes in the posterior distributions.
This was the case for the closure test performed there, where the uncertainty of the test point was sufficiently small.
For real experimental data with a larger uncertainty, we will therefore use a method that can deal with multi-modal posterior distributions.
Since the previously tested PTLMC method does not compute the evidence $\mathcal{P}({\bf y}_{\rm exp})$ (this factor is essential for computing the Bayes factor, which we use to compare different training processes in our analysis), we will employ another efficient MCMC sampler routine implemented in the \texttt{pocoMC} package~\cite{karamanis2022accelerating,
karamanis2022pocomc}.
This sampler implements Preconditioned Monte Carlo (PMC), where first normalizing flow is applied to decorrelate the parameters and the distributions, and then adaptive Sequential Monte Carlo (SMC) is applied.

The Bayesian inference analysis in this paper is performed using the STAR and PHOBOS measurements for Au+Au collisions at the RHIC BES program. The details are listed in Table~\ref{tab:training_data}.

\begin{table}[h!]
    \caption{The experimental measurements for Au+Au collisions used in our Bayesian inference analysis. We include identified ${\rm d}N/{\rm d}y$ and $\langle p_{\rm T} \rangle$ from central to 60\% in collision centrality and the charged hadron $v_n\{2\}$ from 0 to 50\% centrality.}
    \label{tab:training_data}
    \begin{tabular}{c|c|c}
        \hline\hline
        $\sqrt{s_{\rm NN}}$~[GeV] & STAR & PHOBOS \\
        \hline
        200 & $\mathrm{d}N/\mathrm{d}y (\pi^+,K^+,\bar{p},p)$~\cite{STAR:2008med} & $\mathrm{d}N_{\rm ch}/\mathrm{d}\eta$~\cite{PHOBOS:2005zhy} \\
            & $\langle p_{\rm T}\rangle (\pi^+,K^+,p,\bar{p}$~\cite{STAR:2008med} & $v_2^{\rm ch}(\eta)$~\cite{PHOBOS:2006dbo} \\
            & $v_{2}^{\rm ch}\lbrace 2\rbrace$~\cite{STAR:2017idk}, $v_{3}^{\rm ch}\lbrace 2\rbrace$~\cite{STAR:2016vqt} \\
        \hline
        19.6 & $\mathrm{d}N/\mathrm{d}y (\pi^+,K^+,p)$~\cite{STAR:2017sal} & $\mathrm{d}N_{\rm ch}/\mathrm{d}\eta$~\cite{PHOBOS:2005zhy} \\
             & $\langle p_{\rm T}\rangle (\pi^+,K^+,p,\bar{p})$~\cite{STAR:2017sal} \\
             & $v_{2}^{\rm ch}\lbrace 2\rbrace$~\cite{STAR:2017idk}, $v_{3}^{\rm ch}\lbrace 2\rbrace$~\cite{STAR:2016vqt} \\
        \hline
        7.7 & $\mathrm{d}N/\mathrm{d}y (\pi^+,K^+,p)$~\cite{STAR:2017sal} & \\
            & $\langle p_{\rm T}\rangle (\pi^+,K^+,p,\bar{p})$~\cite{STAR:2017sal} & \\
            & $v_{2}^{\rm ch}\lbrace 2\rbrace$~\cite{STAR:2017idk}, $v_{3}^{\rm ch}\lbrace 2\rbrace$~\cite{STAR:2016vqt} \\
        \hline\hline
    \end{tabular}
\end{table}

In Ref.~\cite{Roch:2024xhh}, we systematically studied the performance of different model emulation setups on these experimental measurements and quantified their impacts on the closure test.

\section{Posterior Analysis}
\label{sec:posterior}

In this section, we perform Bayesian analyses on the RHIC measurements using three different GP models: PCSK and Scikit GP trained with LHD + HPP simulations, and PCSK trained only with the LHD dataset.
We will first evaluate and compare the general aspects of the posterior distributions from the three GP models and then discuss the physics insights from the constraints on model parameters.

\begin{figure*}[t!]
    \includegraphics[width=\textwidth]{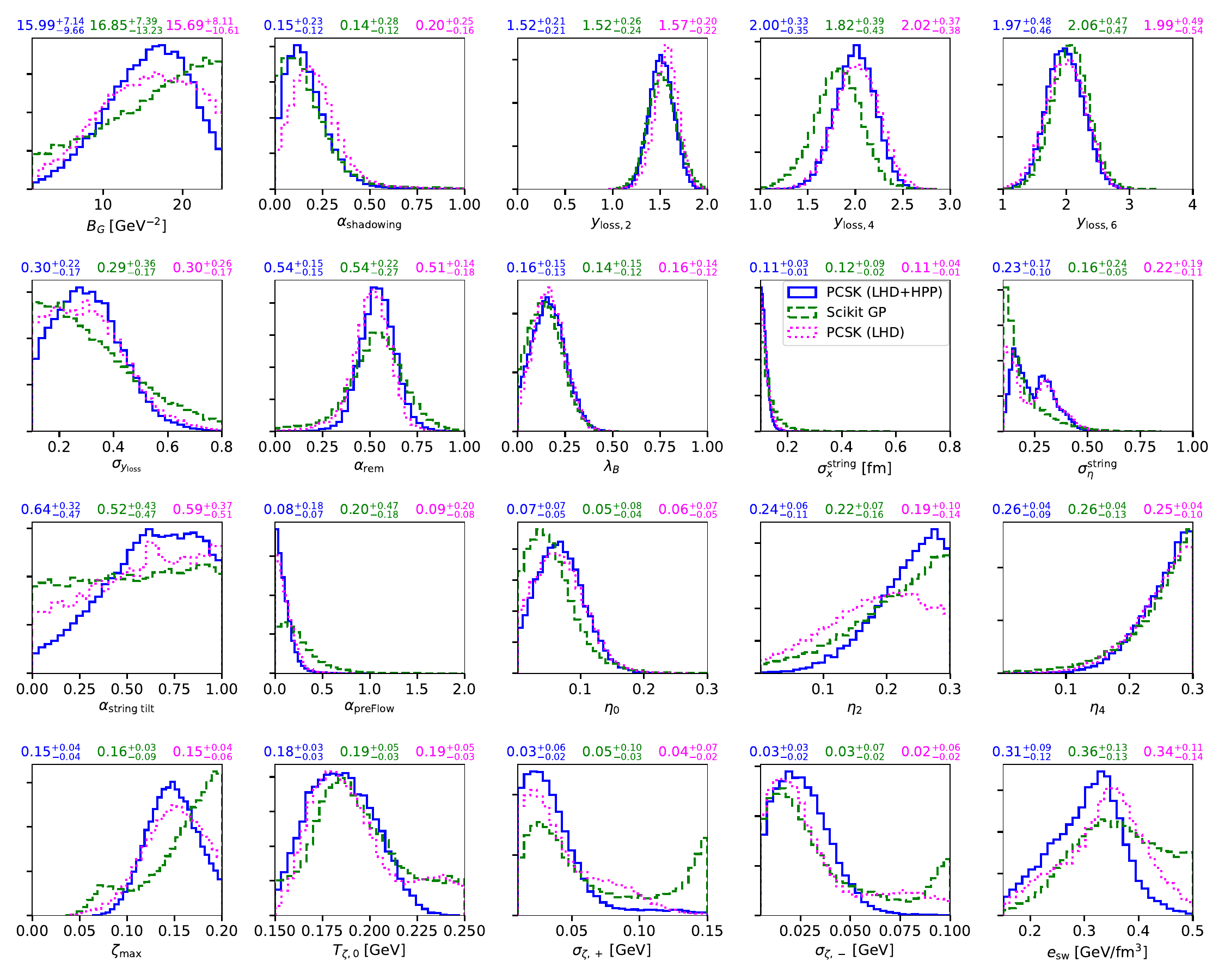}
    \caption{Comparison of the marginal posterior distributions of the 20 model parameters using the MCMC sampling with the PCSK and Scikit GP emulators.
    Values above the plots show the parameters' median and $90\%$ confidence intervals. The $x$-axes represent the ranges of the uniform prior distributions.}
    \label{fig:posterior_comparison}
\end{figure*}

In Fig.~\ref{fig:posterior_comparison}, we show the posterior marginal distributions for the 20 model parameters using the PCSK emulator (blue), the Scikit GP emulator (green), and the PCSK emulator trained only with the 1,000 LHD points (magenta). The former two GP emulators were trained with the 1,095 LHD + HPP training points.
We showed that the PCSK emulator performed much better than the Scikit GP in the closure test in Ref.~\cite{Roch:2024xhh}. However, when applying the Bayesian inference analysis to the real experimental measurements, the two emulators result in similar posterior distributions for the model parameters. 
For most of the parameters, we find that the PCSK emulator still gives slightly more peaked distributions than those from the Scikit GP.

Comparing the blue and magenta posterior distributions, we find that including the additional 95 HPP points in training the GP emulator results in slightly stronger constraints on the model parameters. 

The amount of information gained in our Bayesian inference analysis can be quantified using the Kullback–Leibler (KL) divergence, which is defined as,
\begin{equation}
    D_\mathrm{KL}(p \vert\vert q) = \int{\rm d}\bs{\theta}\; p(\bs{\theta}) \ln\left(\frac{p(\bs{\theta})}{q(\bs{\theta})} \right).
    \label{eq:KLdivergence}
\end{equation}
Here, we compute the KL divergence with $p(\bs\theta)$ as the posterior distribution $\mathcal{P}(\bs{\theta} | {\bf y}_{\rm exp})$ in Eq.~\eqref{eq:bayes} and $q(\bs\theta)$ as the uniform prior distribution $\mathcal{P}(\bs{\theta})$. 
Using Eq.~\eqref{eq:bayes}, we can compute the KL divergence from the sampled posterior chains as
\begin{equation}
    D_\mathrm{KL}(p \vert\vert q) = \frac{1}{N_\mathrm{samples}} \sum_{i} \ln \left(\frac{\mathcal{P}({\bf y}_{\rm exp} | \bs{\theta}_i)}{\mathcal{P}({\bf y}_{\rm exp})} \right),
\end{equation}
where the index $i$ runs through the $N_\mathrm{samples}$ parameter samples in the posterior chains.

\begin{table}[h!]
    \caption{The Kullback-Leibler divergence of posterior distributions to the uniform prior in our Bayesian inference analysis using different GP emulators.}
    \label{tab:KLdivergence}
    \centering
    \begin{tabular}{c|c}
        \hline\hline
        GP Model & $D_\mathrm{KL}$ \\
        \hline
        PCSK (LHD + HPP) & 24.6 \\
        Scikit GP & 20.9 \\
        PCSK (LHD only) & 22.4 \\ 
        \hline\hline
    \end{tabular}
\end{table}

Table~\ref{tab:KLdivergence} reports the KL divergence for the three posterior distributions presented in Fig.~\ref{fig:posterior_comparison}. The ordering of the $D_\mathrm{KL}$ is consistent with our observation in Fig.~\ref{fig:posterior_comparison} discussed above. The more accurate model emulator, PCSK, gives stronger constraints on model parameters than those from the Scikit GP. Including 95 high probability posterior (HPP) points in training the GP emulator further tightens the model constraints.

Another metric to compare the posterior distributions is the Bayes factor. 
The Bayes factor is defined as the ratio of posterior probabilities from two models, $A$ and $B$, given the experimental data $\mathbf{y}_{\rm exp}$,
\begin{align}
    \mathcal{B}_{A/B} \equiv \frac{\mathcal{P}(A \lvert \mathbf{y}_{\rm exp})}{\mathcal{P}(B \lvert \mathbf{y}_{\rm exp})} = \frac{\mathcal{P}(\mathbf{y}_{\rm exp}\lvert A)\mathcal{P}(A)}{\mathcal{P}(\mathbf{y}_{\rm exp}\lvert B)\mathcal{P}(B)}.
    \label{eq:B_factor}
\end{align}
Here, we compare the three GP emulator models shown in Fig.~\ref{fig:posterior_comparison}. Because the three GP models are emulating the same theoretical model, we assume the prior model probability is the same for all the models, namely $\mathcal{P}(A) = \mathcal{P}(B)$ in Eq.~\eqref{eq:B_factor}. Thus, the Bayes factor reduces to the ratio of marginalized likelihoods, also known as Bayesian evidence:
\begin{equation}
    \mathcal{B}_{A/B} = \frac{\mathcal{P}(\mathbf{y}_{\rm exp}\lvert A)}{\mathcal{P}(\mathbf{y}_{\rm exp}\lvert B)}.
    \label{eq:B_evidence}
\end{equation}
The Bayesian evidence for a model $A$ on the given experimental data $\mathbf{y}_{\rm exp}$, $\mathcal{P}(\mathbf{y}_{\rm exp}\lvert A)$, can be computed by integrating the likelihood function over the model parameter set across the entire parameter space:
\begin{equation}
    \mathcal{P}(\mathbf{y}_{\rm exp}\lvert A) = \int\mathrm{d}{\bs\theta}_{A}\;\mathcal{P}(\mathbf{y}_{\rm exp}\lvert {\bs\theta}_{A}, A)\mathcal{P}({\bs\theta}_{A}).
    \label{eq:B_marginalized_L}
\end{equation}
We compute the Bayesian evidence for the three models using the \texttt{pocoMC} package. The Bayes factors are reported in Table~\ref{tab:Bayes_factor}.

Following the Jeffreys' Scale of the Bayes factor in Ref.~\cite{doi:10.1080/00107510802066753}, we find strong evidence in favor of the analysis with the PCSK emulator than the one with the Scikit GP. This result emphasizes the importance of getting an accurate model emulator in Bayesian studies.
We also compare the two model setups with different training data sets using the PCSK emulator. In this case, we find a Bayes factor with a magnitude between 1 and 2.5, which suggests that the PCSK trained with only the LHD points is favored but only with weak evidence. 
Naively, one would expect the GP emulator with more training data to be more favored. The reversed hierarchy here is because the PCSK emulator trained with the LHD + HPP dataset gave a smaller model covariance matrix $\bs{\Sigma}_\mathrm{model}$, which leads to an overall lower likelihood in Eq.~\eqref{eq:likelihood} compared to the results from PCSK trained with LHD only.

Considering the KL divergence and Bayes factor results, we will perform further analysis on the posterior distribution from the PCSK emulator (LHD + HPP) from now on.

\begin{table}[h!]
    \caption{Bayes factors for different combinations in the model setup.}
    \label{tab:Bayes_factor}
    \centering
    \begin{tabular}{c|c|c}
        \hline\hline
        Model $A$ & Model $B$ & $\ln(\mathcal{B}_{A/B})$ \\
        \hline
        PCSK (LHD + HPP) & Scikit GP & $6.94\pm 0.03$ \\
        PCSK (LHD + HPP) & PCSK (LHD only) & $-1.41\pm 0.03$ \\
        \hline\hline
    \end{tabular}
\end{table}

\begin{figure}[h!]
    \centering
    \includegraphics[width=0.9\linewidth]{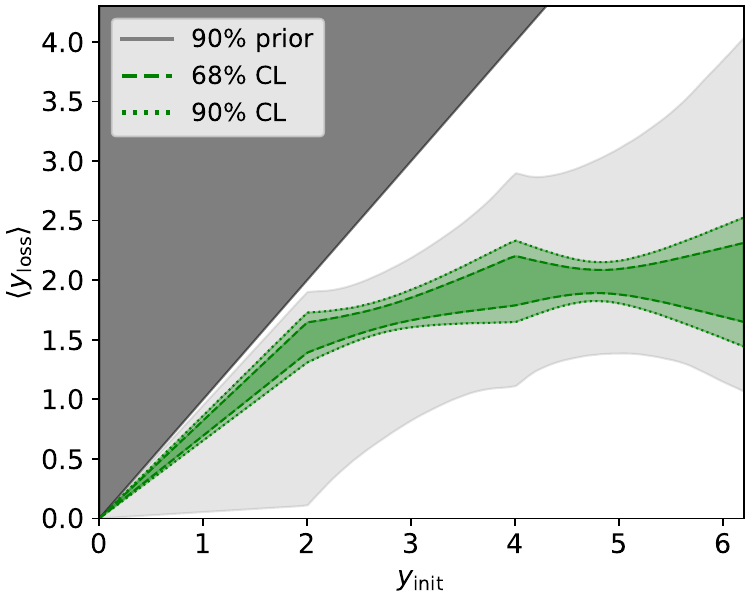}
    \includegraphics[width=0.95\linewidth]{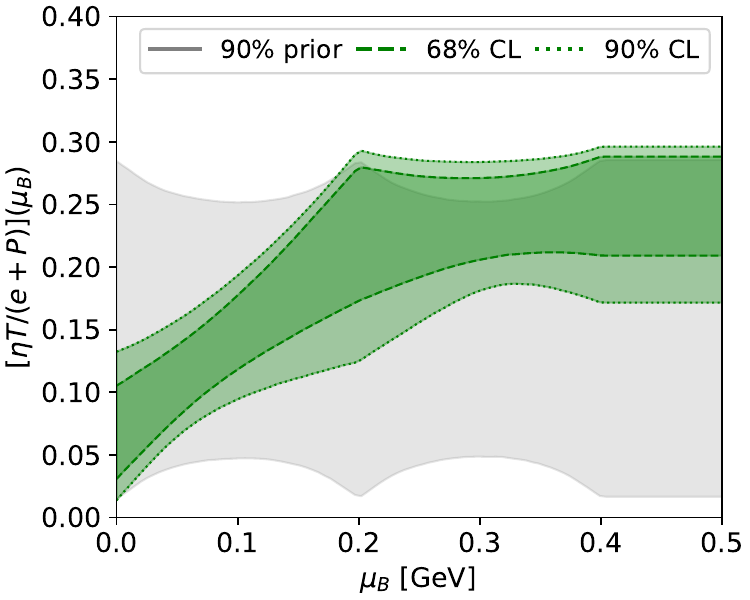}
    \includegraphics[width=0.95\linewidth]{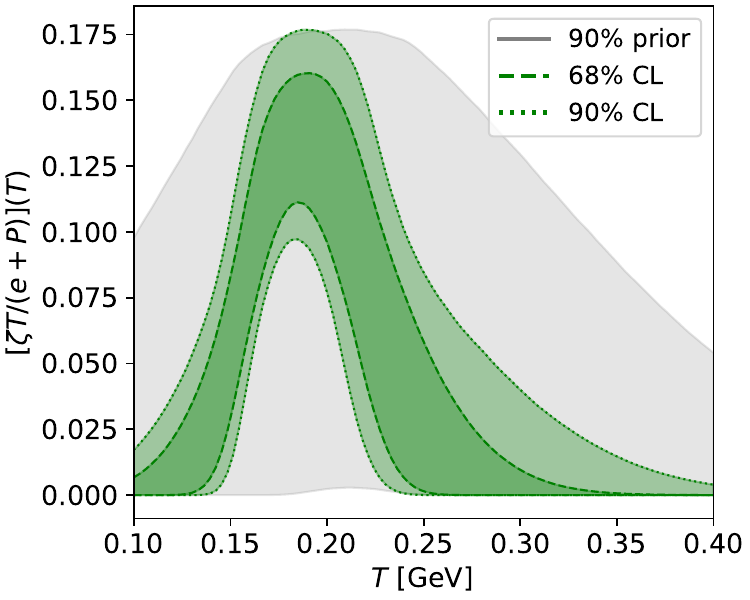}
    \caption{Posterior distributions for the average rapidity loss $\langle y_{\rm loss}\rangle$ (top) as a function of the initial rapidity, the QGP specific shear viscosity (middle) and bulk viscosity (bottom) as functions of $\mu_B$ and temperature respectively. Bands indicate the 68\% (dashed) and 90\% (dotted) confidence levels. The 90\% prior is shown as a gray band. For all curves, the PCSK (LHD + HPP) emulator is used in the MCMC.}
    \label{fig:posterior_parametrizations}
\end{figure}

Now, we discuss the physics implications of the constraints on model parameters from the posterior distribution. 

\begin{itemize}
    \item The parameter $B_G$ is the width of the nucleon to sample its hot spots~\cite{Shen:2022oyg}. The posterior range $B_G = 15.99^{+7.14}_{-9.66}$ GeV$^{-2}$ translates to the nucleon mass width between 0.5 and 0.95 fm at 90\% confidence level.
    \item The parameter $\alpha_\mathrm{shadowing}$ controls the probability of producing strings from secondary binary collisions in the \texttt{3D-Glauber} initial state model~\cite{Shen:2017bsr, Shen:2022oyg}. Our Bayesian analysis prefers $\alpha_\mathrm{shadowing} \sim 0.15$, indicating that only 15\% of the binary collisions are shadowed for soft string production. The numbers of produced strings scale closer to the number of binary collisions than the number of participants.
    \item The rapidity loss fluctuation parameter $\sigma_{y_\mathrm{loss}} \sim 0.3$ in our analysis, which is smaller than the value $\sigma_{y_\mathrm{loss}} = 0.6$ used in Ref.~\cite{Shen:2022oyg}. The latter was calibrated using the multiplicity distribution in $p$+$p$ collisions, which has a stronger sensitivity to this model parameter than the 10\% centrality binned particle yields in Au+Au collisions~\cite{Shen:2022oyg}. The particle multiplicity fluctuations in Au+Au collisions include the number of participant fluctuations, further diluting the sensitivity to the rapidity loss fluctuation parameter.
    \item The parameter for the nucleons' remnant rapidity loss $\alpha_\mathrm{rem} = 0.54 \pm 0.15$ is constrained by the shape of the pseudo-rapidity distribution ${\rm d}N^\mathrm{ch}/{\rm d}\eta$. Its value is consistent with the previous simulations~\cite{Shen:2022oyg}.
    \item The parameter $\lambda_B$ controls the amount of initial-state baryon charge transport to the string junction~\cite{Shen:2022oyg}. Its constrained value $\lambda_B \sim 0.15$ is consistent with previous analyses~\cite{Shen:2022oyg, Pihan:2024lxw}.
    \item Our Bayesian analysis prefers $\alpha_\mathrm{preflow} \sim 0$, suggesting small pre-equilibrium flow in the transverse plane for the string source terms before depositing into hydrodynamics.
    \item The posterior distribution for the hotspot transverse size $\sigma^\mathrm{string}_x$ peaks at the lower bound of the prior, suggesting a small hotspot size is preferred. 
    \item The parameter $\sigma^\mathrm{string}_\eta$ controls the string profile's fall-off along the longitudinal direction. Its marginal distribution has a double-peaked structure, indicating this parameter may have a $\sqrt{s_{\rm NN}}$ dependence.
\end{itemize}

To quantify the constraints on the functional dependence for the initial-state rapidity loss and the QGP-specific viscosities, we display the 90\% confidence level of the prior distributions (gray) with the 68\% and 90\% confidence levels of the posterior distributions (green) for these parameters in Fig.~\ref{fig:posterior_parametrizations}.

The average rapidity loss in the initial state is, compared to the prior distribution (gray), tightly constrained by the particle yield measurements at multiple collision energies. Despite using different types of GP emulators and MCMC algorithms, our results are consistent with those in Ref.~\cite{Shen:2023awv}.
We also obtain strong constraints on the specific shear and bulk viscosities of QGP as functions of $\mu_B$ and $T$, respectively.
The values of QGP-specific shear viscosity show a significant increase with $\mu_B$ within [0, 0.2] GeV. This constraint primarily comes from the pseudo-rapidity dependence of elliptic flow $v_2(\eta)$ measurements~\cite{Shen:2023pgb}.
The values of $\tilde{\eta}$ at larger $\mu_B$ remain around 0.3 closer to the upper bound of the prior for the parameter $\eta_4$.
For the QGP's specific bulk viscosity, we find a peaked distribution around a temperature of $\approx 200\;\mathrm{MeV}$ with a maximum value of 0.12. The values of $\tilde{\zeta}$ fall off to zero quickly when departing from the maximum value.

\section{Model Simulations with Posterior Distributions}
\label{sec:model_sim}
\begin{figure*}[t!]
    \centering
    \includegraphics[width=0.9\textwidth]{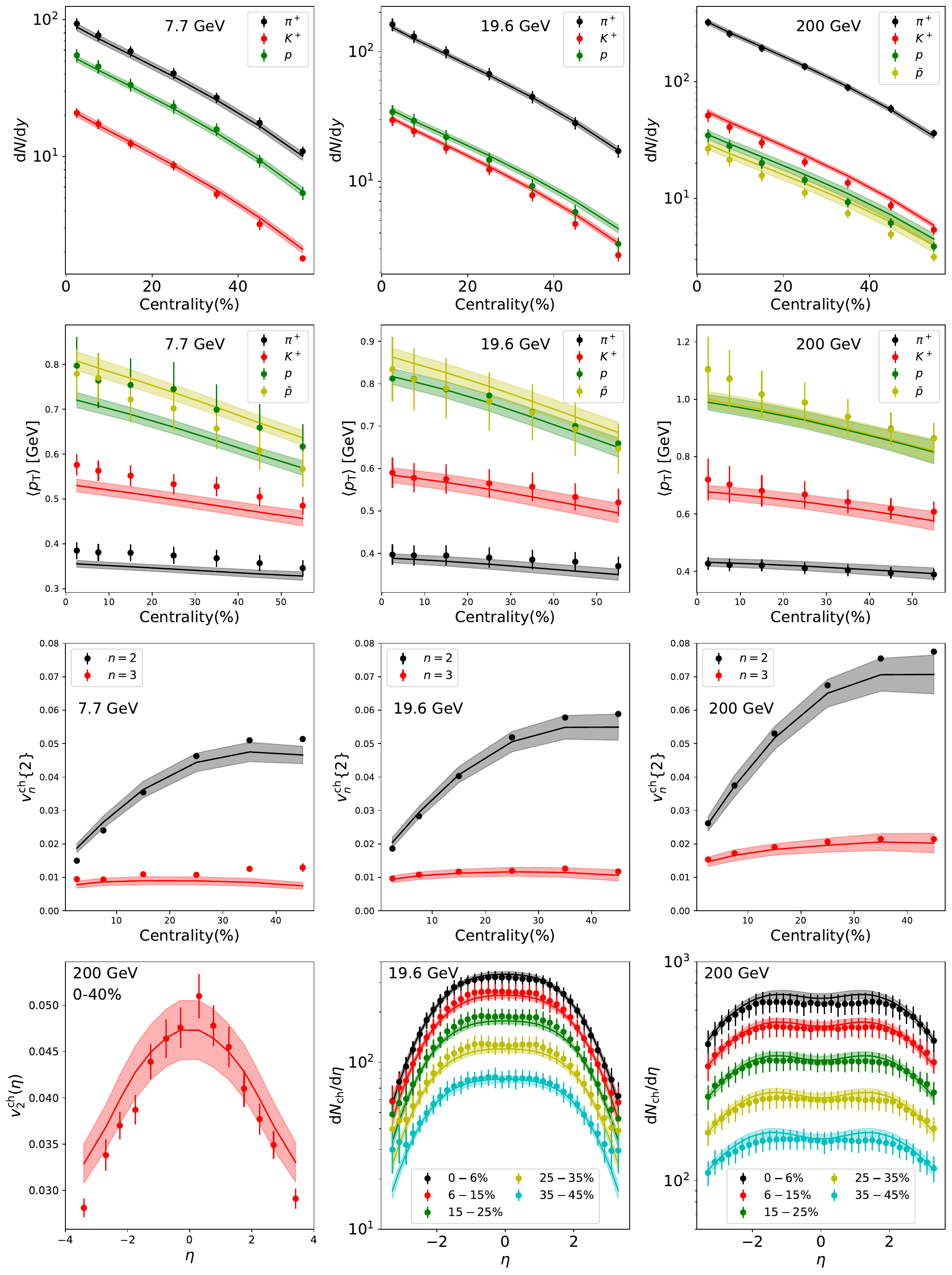}
    \caption{Observables computed with the full theoretical model at 100 posterior samples $\{\bs{\theta}^\mathrm{post}\}$. For the three different collision energies $\sqrt{s_{\rm NN}}=7.7$~GeV, 19.6~GeV, and 200~GeV, we simulate 3,000, 2,000, and 1,000 minimum bias Au+Au collisions at each posterior point $\bs{\theta}^\mathrm{post}_i$, respectively. The full line indicates the mean value of the observable averaged from the 100 parameter sets, and the band indicates the one standard deviation interval. The experimental data are summarized in Table~\ref{tab:training_data}.}
    \label{fig:observables}
\end{figure*}

After using the GP emulators to obtain the posterior distributions of model parameters displayed in Fig.~\ref{fig:posterior_comparison}, we will verify these constraints with full model simulations and further make model predictions. We randomly choose 100 samples $\{\bs{\theta}^\mathrm{post}\}$ from the posterior distributions generated with the PCSK emulator (LHD + HPP) and then perform event-by-event simulations with the \texttt{iEBE-MUSIC} framework.
At each posterior model parameter set $\bs{\theta}^\mathrm{post}_i$, we simulate 1,000 minimum bias Au+Au collisions at $\sqrt{s_{\rm NN}}=200\;\mathrm{GeV}$, 2,000 events at $\sqrt{s_{\rm NN}}=19.6\;\mathrm{GeV}$, and 3,000 collisions at $\sqrt{s_{\rm NN}}=7.7\;\mathrm{GeV}$.
With these simulations, we compute the averages of the observables from the 100 posterior parameter sets and denote their standard deviations as the remaining systematic uncertainties after the Bayesian analysis.
Here, we assume the statistical errors of observables at individual posterior parameter sets are much smaller compared to the standard deviations of the 100 posterior parameter sets.
In Fig.~\ref{fig:observables}, we show the comparison to the experimental observables (see Tab.~\ref{tab:training_data}) used in the Bayesian inference.

\begin{figure*}[t!]
    \centering
    \includegraphics[width=0.32\textwidth]{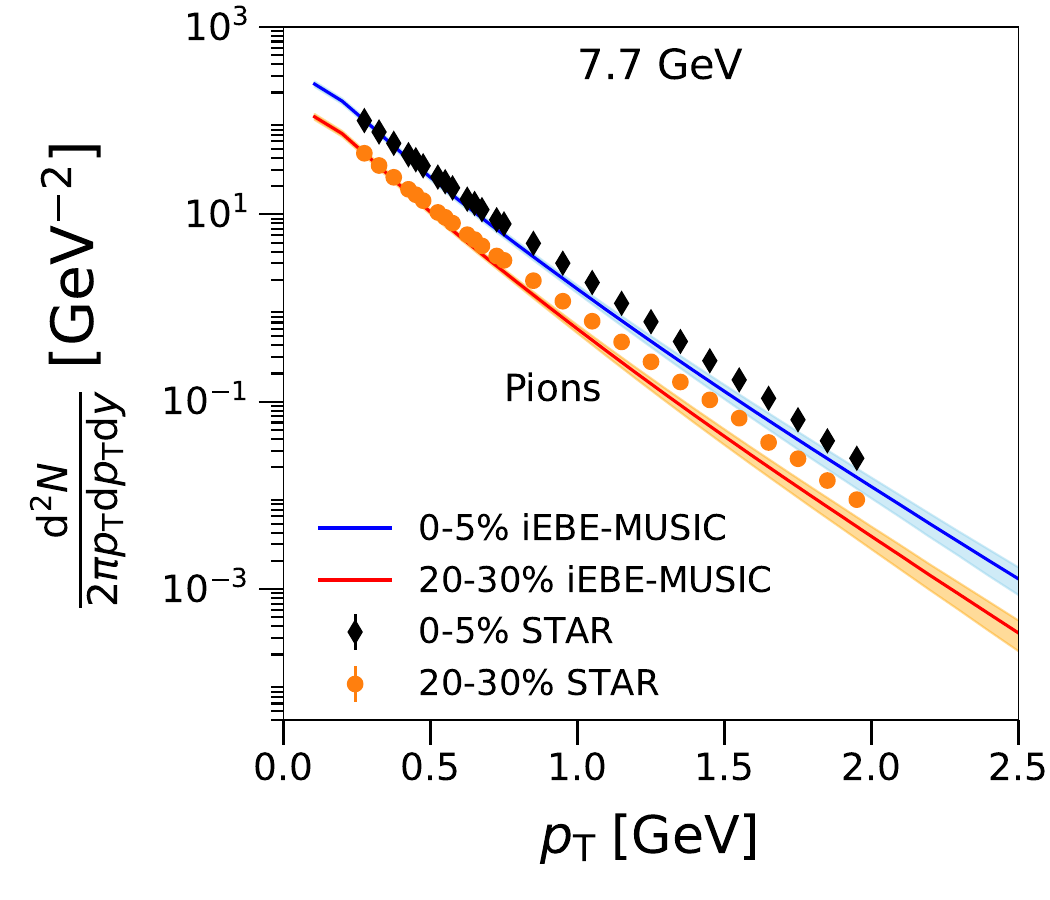} 
    \includegraphics[width=0.32\textwidth]{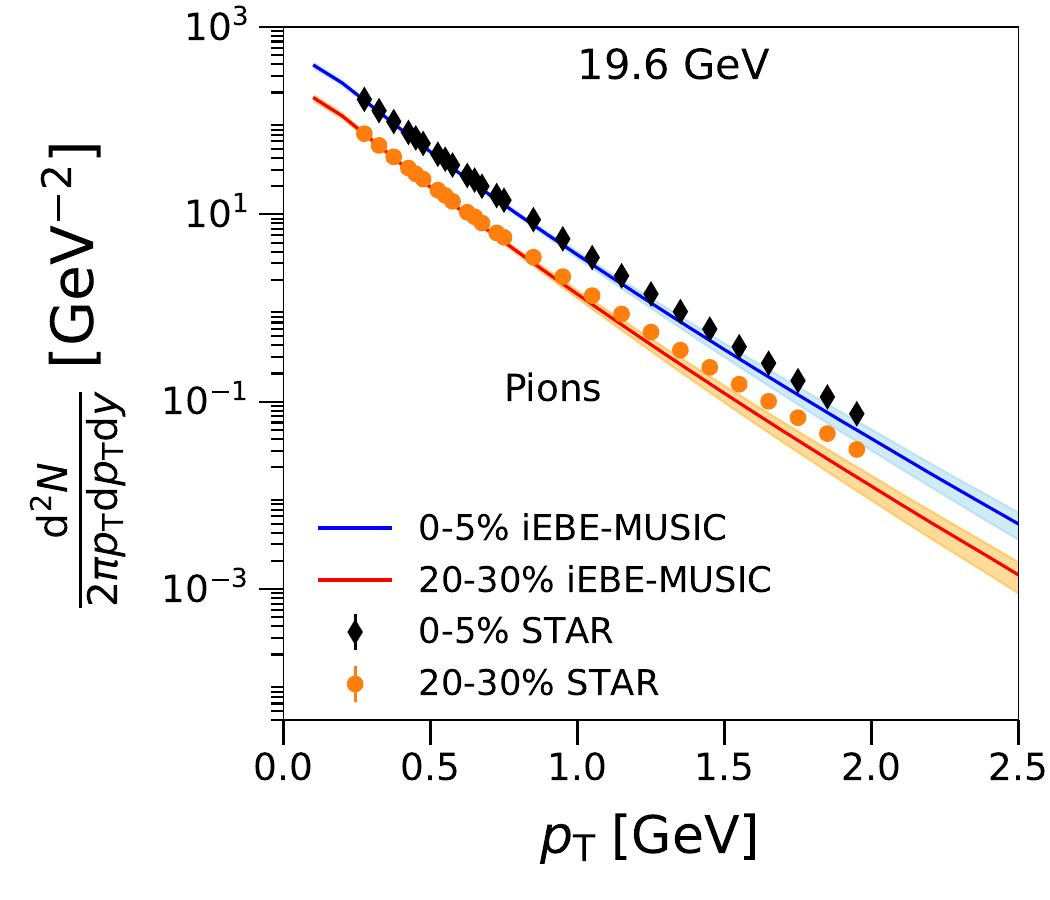} 
    \includegraphics[width=0.32\textwidth]{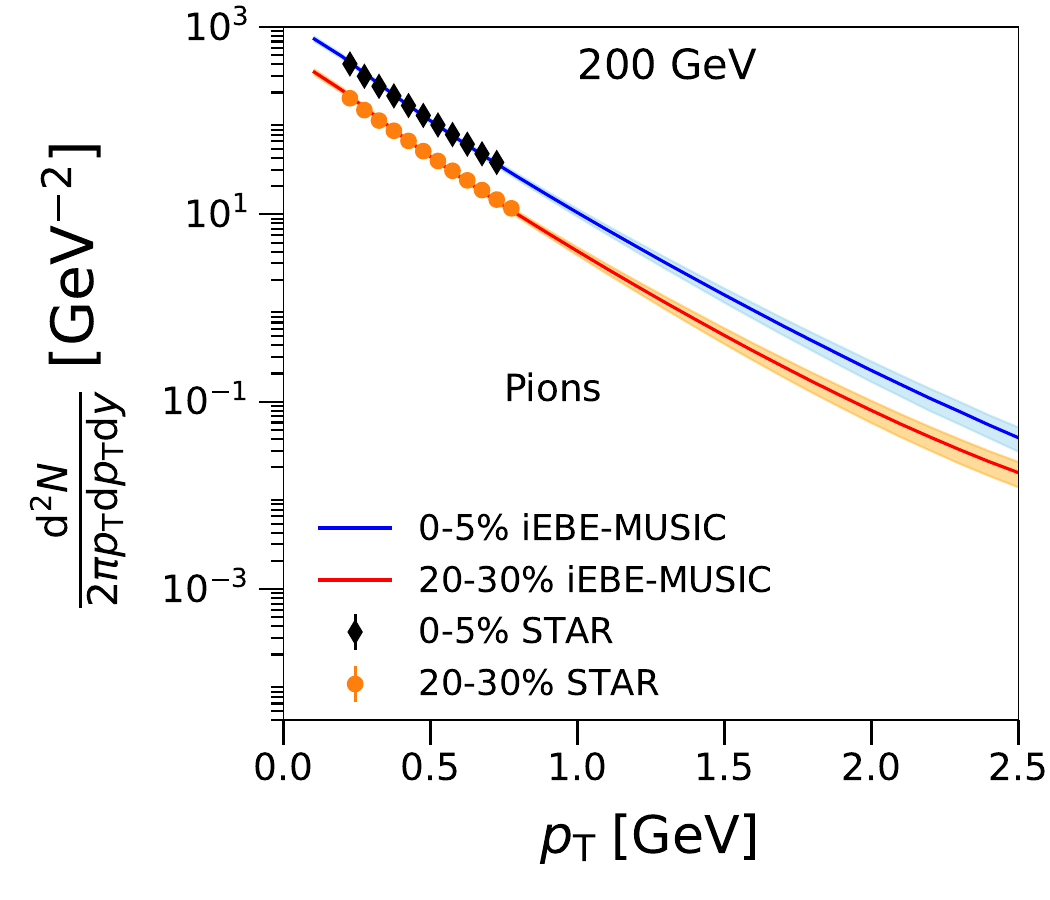}
    \includegraphics[width=0.32\textwidth]{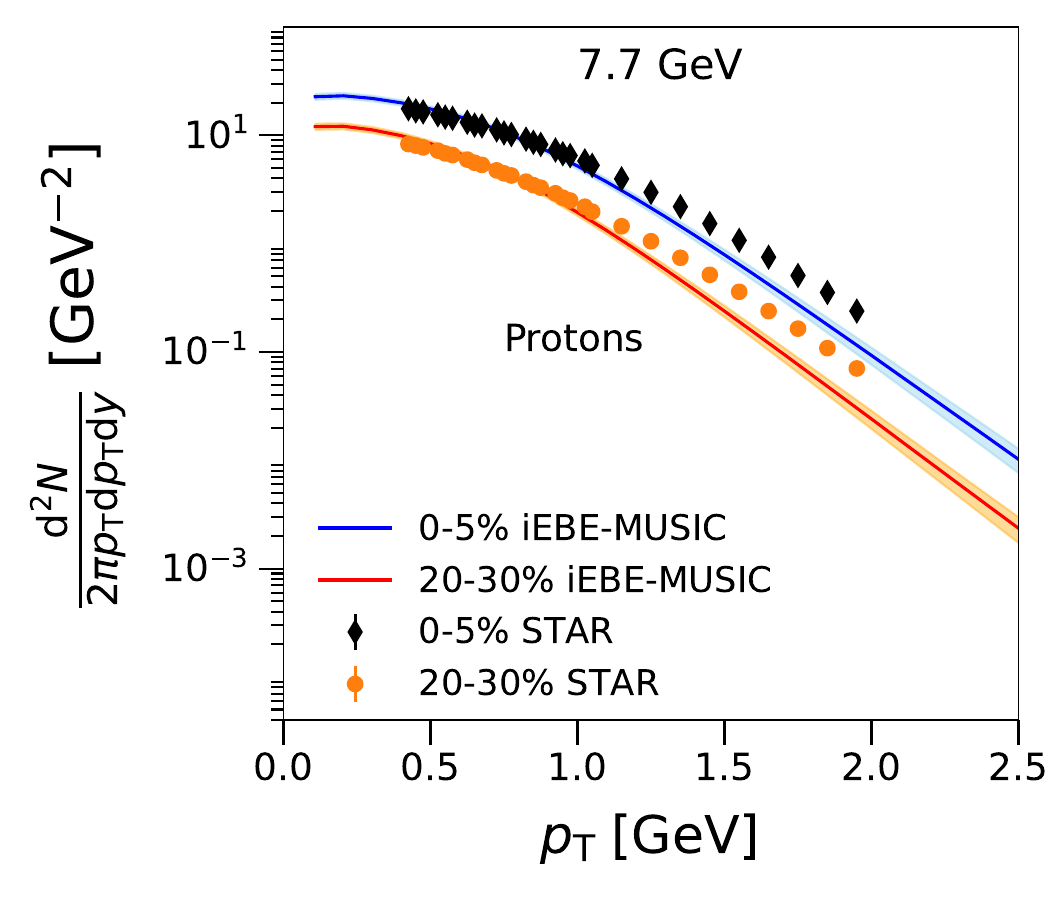}
    \includegraphics[width=0.32\textwidth]{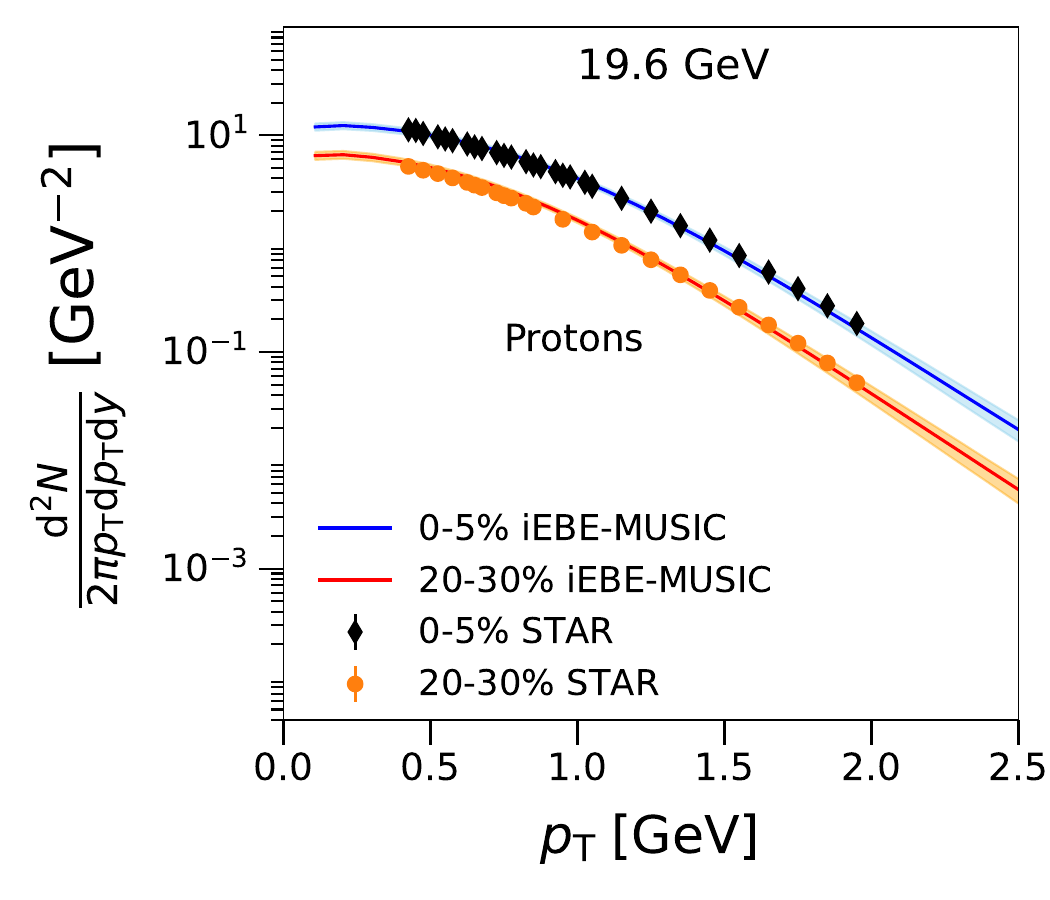}
    \includegraphics[width=0.32\textwidth]{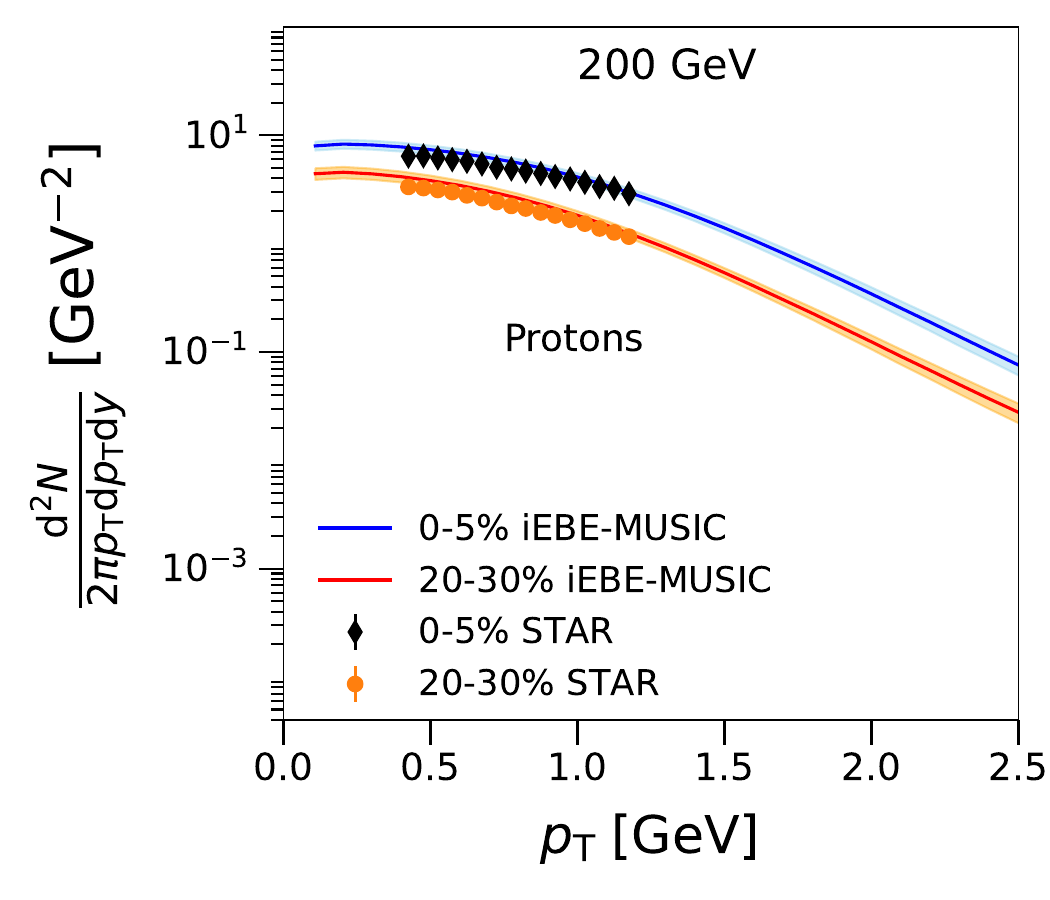}
    \includegraphics[width=0.32\textwidth]{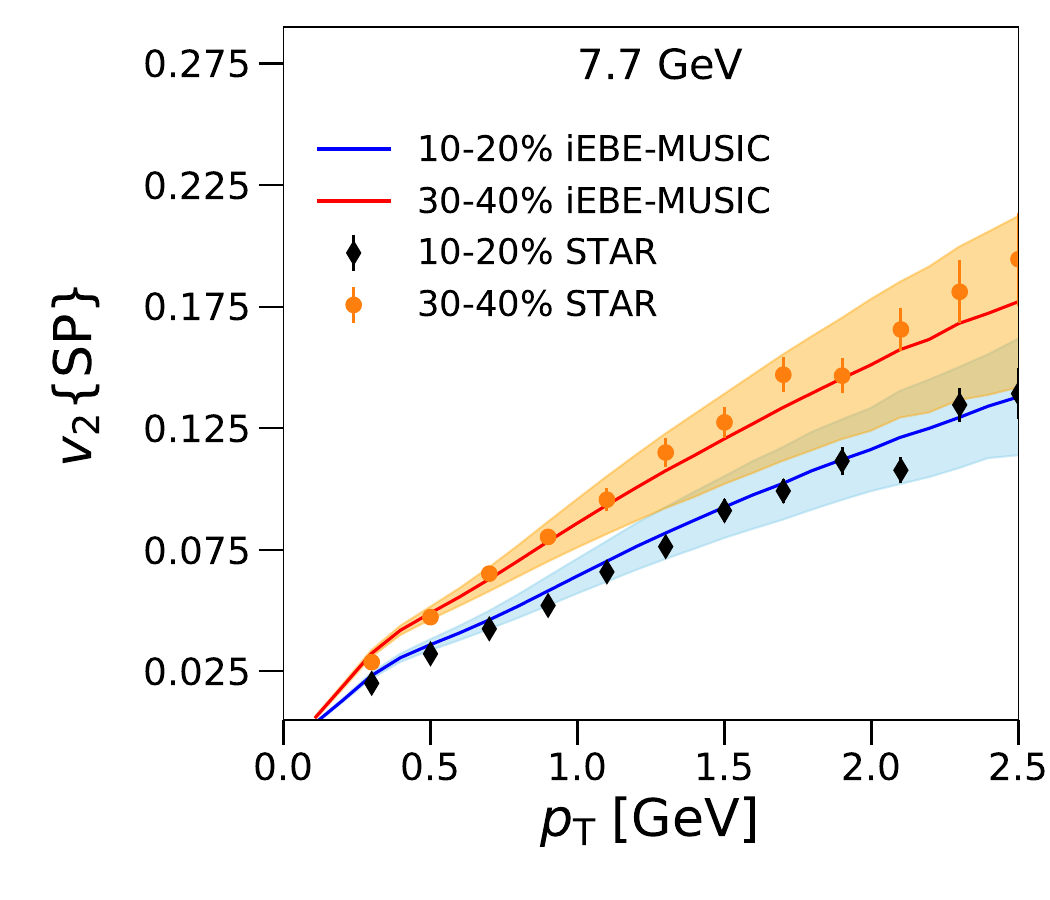} 
    \includegraphics[width=0.32\textwidth]{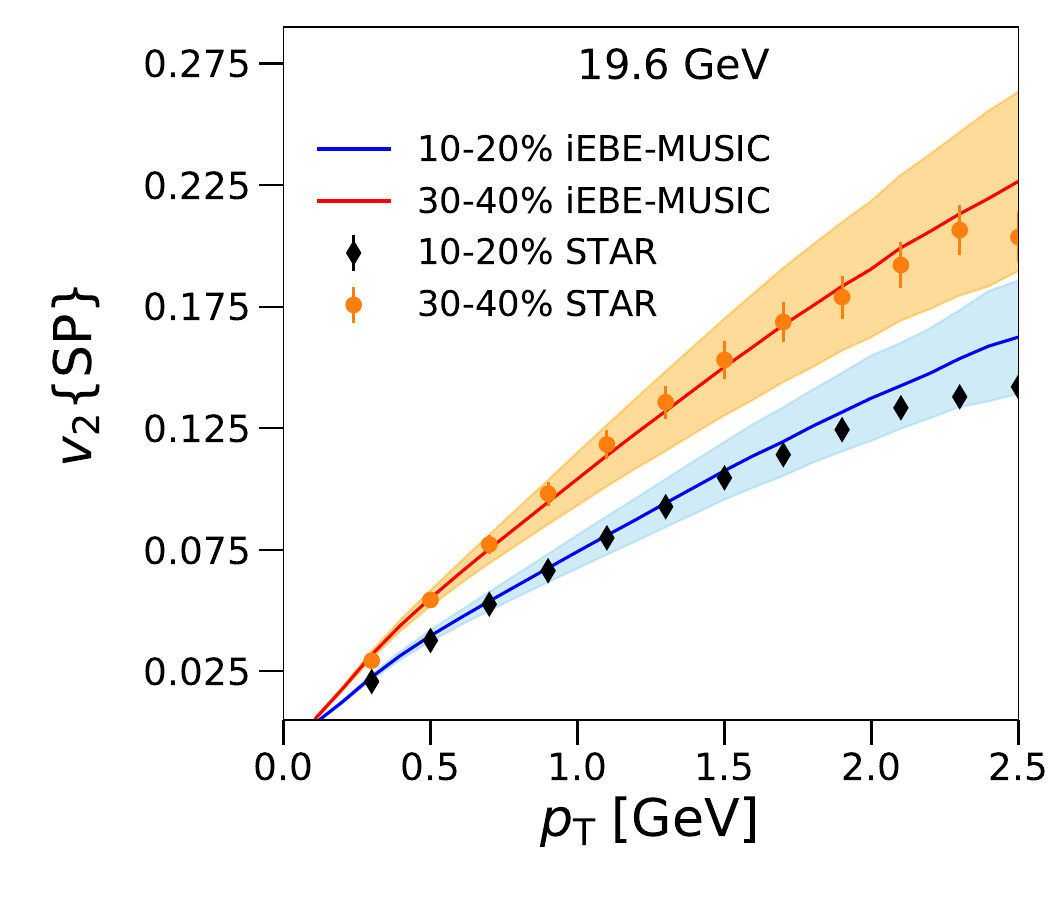}
    \includegraphics[width=0.32\textwidth]{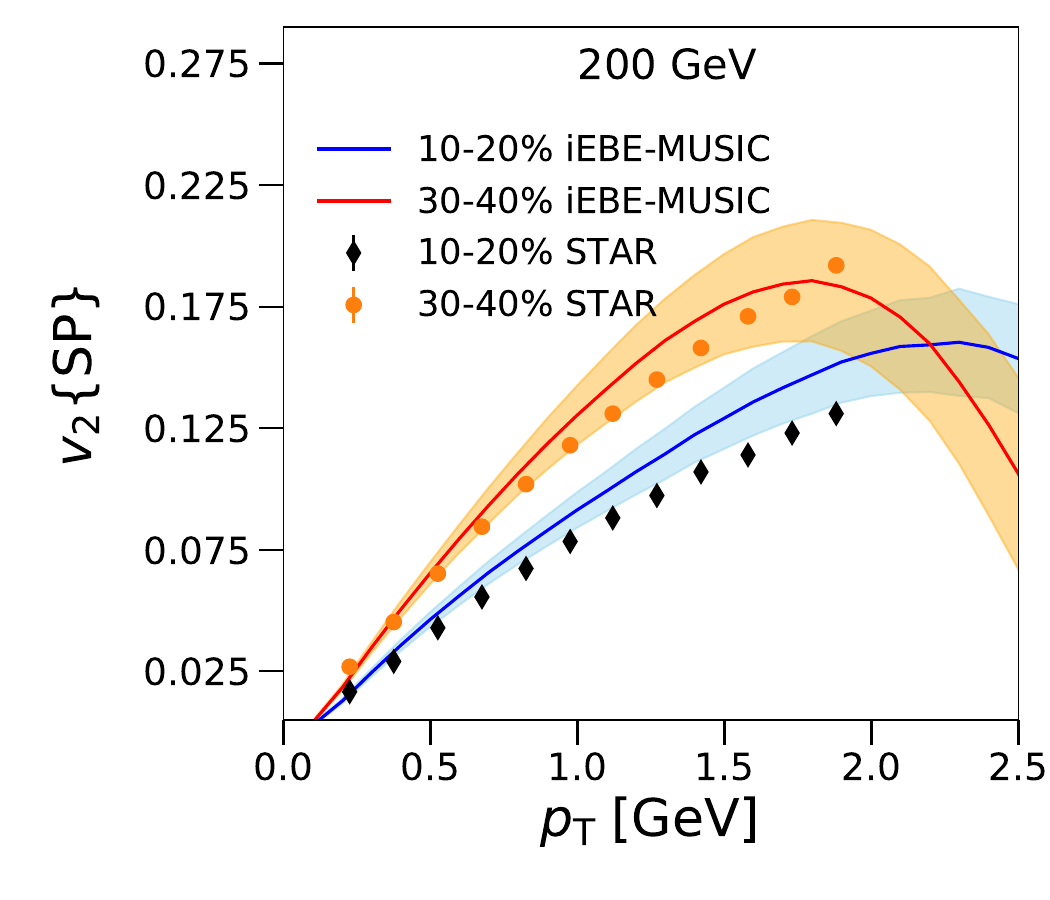} 
    \caption{Model predictions for $p_{\rm T}$-differential observables from the 100 posterior parameter samples $\{\bs\theta^\mathrm{post}\}$.
    The three columns from left to right represent observables at $\sqrt{s_{\rm NN}}=7.7$~GeV, $19.6$~GeV, and $200$~GeV, respectively.
    The first two rows show $p_{\rm T}$-differential $\pi^+$ and $p$ spectra compared with the STAR measurements~\cite{STAR:2017sal, STAR:2008med}, while the bottom row shows the $p_{\rm T}$-differential elliptic flow $v_{2}\{\mathrm{SP}\}$ compared with the STAR data~\cite{STAR:2012och, STAR:2008ftz}. The band indicates the one standard deviation interval computed from the 100 parameter sets. }
    \label{fig:model_predictions}
\end{figure*}

In general, the full model calculations with parameters sampled from the posterior distribution can provide a good description of the 544 data points in the Bayesian inference. The agreement is generally better for central collisions and near the mid-rapidity region. This behavior is related to the fact that the GP emulator has smaller emulation errors for those observables than the observables in peripheral centralities and forward rapidity regions~\cite{Roch:2024xhh}. 

Looking into details of the mid-rapidity observables at $\sqrt{s_{\rm NN}}=200$~GeV (right columns of Fig.~\ref{fig:observables}), our model slightly overpredicted the particle yields of $K^+$ mesons. The centrality dependencies of proton and anti-proton yields are somewhat flatter than the STAR measurements. All these observables are sensitive to the switching energy density from hydrodynamics to hadronic transport, the current model discrepancy could be reduced by allowing the parameter $e_\mathrm{sw}$ to be centrality dependent.
The identified particles' mean $p_{\rm T}$ and mass hierarchy reflect the amount of radial flow developed in the hydrodynamic phase. Our model provides a good description of the STAR measurements for pions and kaons, while slightly underestimating the proton and anti-proton mean $p_{\rm T}$ but still within the experimental uncertainties. The averaged transverse momenta of protons is more sensitive to the scattering cross sections and baryon-anti-baryon annihilation processes in the hadronic transport model~\cite{Ryu:2017qzn}.

For the mid-rapidity observables at $\sqrt{s_{\rm NN}}=19.6$~GeV, the description of kaon's particle yields is slightly improved compared to that at 200 GeV. At this collision energy, the averaged net baryon chemical potential can reach up to 150-200 MeV at mid-rapidity~\cite{STAR:2017sal}. Our model gives a $\sim 50$~MeV larger mean $p_{\rm T}$ for anti-protons than those for protons at finite baryon density. The current STAR measurement can not distinguish this difference. In our model, the mean $p_{\rm T}$ difference between protons and anti-protons primarily comes from the \texttt{UrQMD} hadronic scattering phase as anti-protons emit relatively earlier than protons from the hydrodynamic hypersurface and experience more hadronic interactions~\cite{Denicol:2018wdp}. This phenomenon is more pronounced at $\sqrt{s_{\rm NN}}=7.7$~GeV, at which the anti-proton mean $p_{\rm T}$ is about $100$~MeV larger than that of protons in our calculations. At $\sqrt{s_{\rm NN}}=7.7$~GeV, our model underestimates the mean $p_{\rm T}$ of pions and kaons, suggesting the systems do not develop enough radial flow in the hydrodynamic phase. Introducing an explicit $\mu_B$ dependence on the QGP-specific bulk viscosity would help to reduce the tension here. Because the hydrodynamic phase at $\sqrt{s_{\rm NN}}=7.7$~GeV is limited, we observe that the triangular flow coefficients $v_3^{\rm ch}\{2\}$ start to deviate from the experimental data in peripheral collisions, similar to the finding in Ref.~\cite{Schafer:2021csj}.

Lastly, we look at the pseudorapidity-dependent observables. Our model gives a good description of the PHOBOS charged hadron ${\rm d}N^\mathrm{ch}/{\rm d}\eta$ measurements across collision centrality at 200 and 19.6 GeV.
Our model gives a flatter shape for $v_2^{\rm ch}(\eta)$ compared to the PHOBOS measurements. This is because our parameterization for the QGP shear viscosity is temperature-independent. Introducing a $\eta/s(T)$ will improve the model-to-data comparison~\cite{Denicol:2015nhu}.

The full model simulations allow us to go beyond the calibrated experimental measurements and make model predictions with controlled systematic uncertainties. In this work, we select $p_{\rm T}$-differential observables to demonstrate the predictive power of our model. We will report a systematic extrapolation of our model to different collision systems such as ($p$, $d$, $^3$He)+Au, $O$+$O$, Ru+Ru, Zr+Zr collisions, and Au+Au collisions at other collision energies in the RHIC BES program in future work.

Since our model is calibrated using only the $p_{\rm T}$-integrated observables in Table~\ref{tab:training_data}, the model-to-data comparison for $p_{\rm T}$-differential observables can effectively test the consistency of our Bayesian analysis.

Figure~\ref{fig:model_predictions} shows our model predictions for the $p_{\rm T}$-differential spectra for pions and protons and charged hadron $v_2(p_{\rm T})$ in Au+Au collisions at 7.7, 19.6, and 200 GeV compared with the STAR measurements. We find a good description of the identified particle $p_{\rm T}$ spectra at 200 and 19.6 GeV, while the $p_{\rm T}$ spectra are slightly steeper than the STAR data at 7.7 GeV. These results are consistent with our description of the identified particle mean $p_{\rm T}$ in Fig.~\ref{fig:observables}. Our results show that the first two moments of the particle $p_{\rm T}$ spectra, namely the particle yield and mean $p_{\rm T}$, are sufficient to produce reasonable $p_{\rm T}$ spectra in the Bayesian calibration~\cite{Pratt:2015zsa}.

For the charged hadron $p_{\rm T}$-differential elliptic flow, our model prediction can reproduce the experimental data fairly well up to 2 GeV. We notice that the systematic uncertainty in $v_2(p_{\rm T})$ grows with $p_{\rm T}$ in the calculations. The effects from the out-of-equilibrium $\delta f$ corrections in the Cooper-Frye particlization become important at high $p_{\rm T}$. The $v_2(p_{\rm T})$ at 200 GeV shows a strong suppression in $p_{\rm T} > 2$ GeV from the $\delta f$ based on the Grad's moment method.

\section{High Likelihood Parameters}
\label{sec:high_likelihood_parameters}

Although our approach in Sec.~\ref{sec:posterior} is a preferred way to make model predictions with controlled systematic errors, it is computationally demanding to perform event-by-event simulations at $\mathcal{O}(100)$ parameter sets sampled from the posterior distribution, especially for observables which require high statistics.

\begin{table}[b!]
    \caption{The highest likelihood parameter set obtained from the MCMC with the PCSK emulator (LHD + HPP), imposing the monotonic constraint $y_{\rm loss,2}\leq y_{\rm loss,4}\leq y_{\rm loss,6}$.}
    \label{tab:high_L}
    \centering
    \begin{tabular}{c|c|c|c}
        \hline\hline
        Parameter & Value & Parameter & Value \\
        \hline
        $B_G\;[\mathrm{GeV}^{-2}]$ & 17.095 & $\alpha_{{\rm string}\;{\rm tilt}}$ & 0.884 \\
        $\alpha_{\rm shadowing}$ & 0.145 & $\alpha_{\rm preFlow}$ & 0.004 \\
        $y_{{\rm loss},2}$ & 1.467 & $\eta_0$ & 0.045 \\
        $y_{{\rm loss},4}$ & 1.759 & $\eta_2$ & 0.280 \\
        $y_{{\rm loss},6}$ & 2.260 & $\eta_4$ & 0.287 \\
        $\sigma_{y_{\rm loss}}$ & 0.356 & $\zeta_{\rm max}$ & 0.148 \\
        $\alpha_{\rm rem}$ & 0.611 & $T_{\zeta,0}\;[{\rm GeV}]$ & 0.214 \\
        $\lambda_B$ & 0.129 & $\sigma_{\zeta,+}\;[{\rm GeV}]$ & 0.018 \\
        $\sigma_x^{\rm string}\;[{\rm fm}]$ & 0.113 & $\sigma_{\zeta,-}\;[{\rm GeV}]$ & 0.040 \\
        $\sigma_\eta^{\rm string}$ & 0.156 & $e_{\rm sw}\;[{\rm GeV}/{\rm fm}^3]$ & 0.350 \\
        \hline\hline
    \end{tabular}
\end{table}

\begin{figure}[t!]
    \centering
    \includegraphics[width=0.9\linewidth]{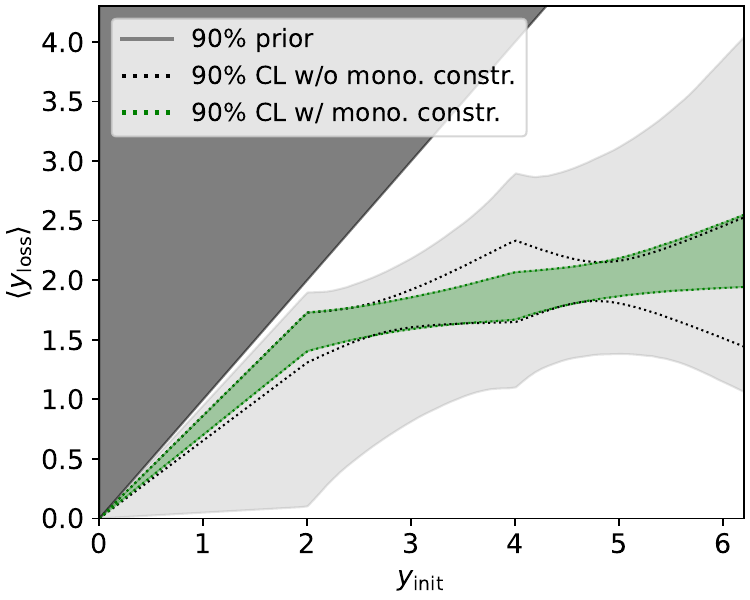}
    \caption{Posterior distributions for the average rapidity loss $\langle y_{\rm loss}\rangle$ as a function of $y_{\rm init}$. The band indicates the 90\% confidence level imposing the monotonic constraint $y_{\rm loss,2}\leq y_{{\rm loss},4}\leq y_{{\rm loss},6}$. The black dotted line shows the 90\% confidence interval without the constraint from Fig.~\ref{fig:posterior_parametrizations}, and the 90\% prior is shown as a gray band.}
    \label{fig:posterior_yloss_constraint}
\end{figure}
\begin{figure}[h!]
    \centering
    \includegraphics[width=\linewidth]{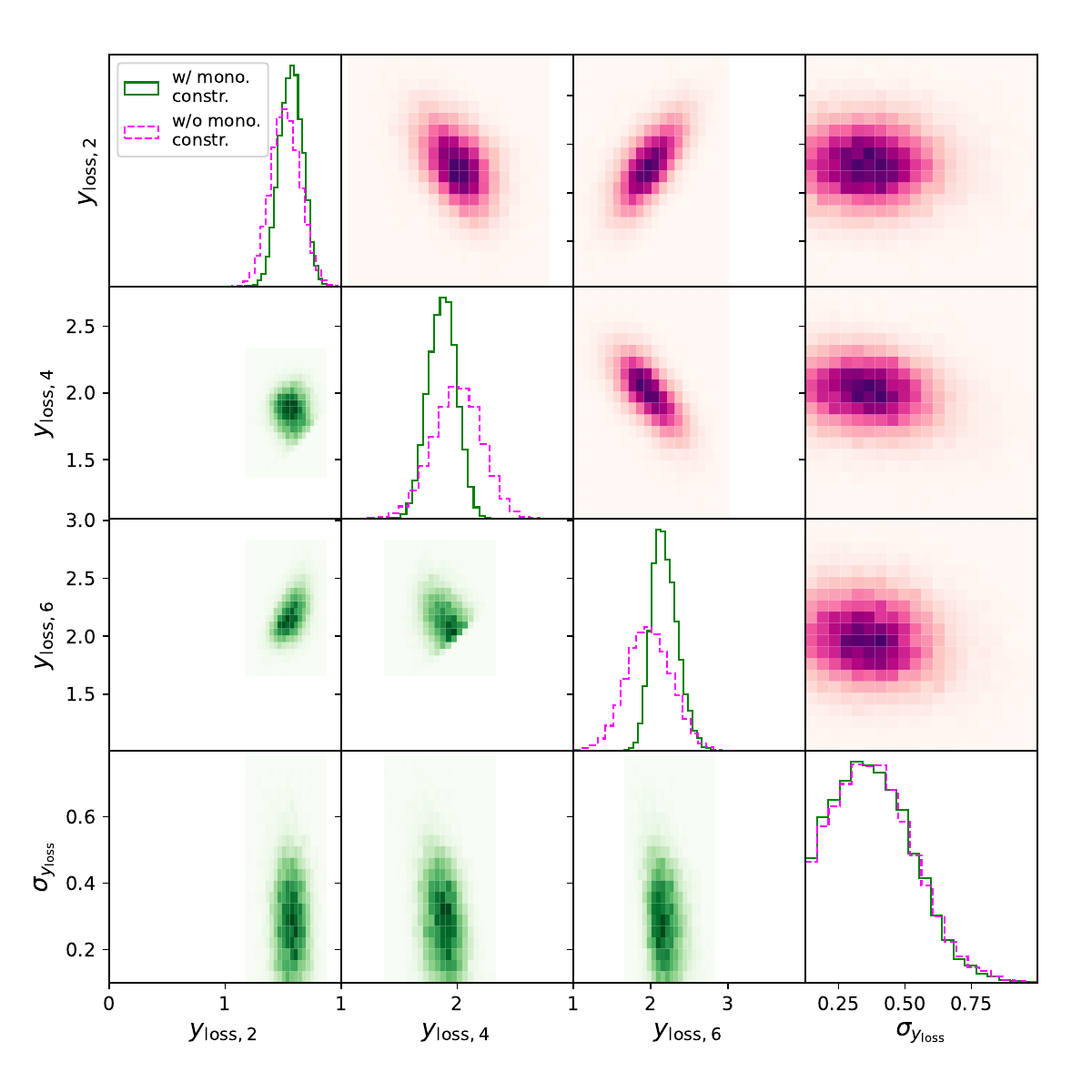}
    \caption{Single parameter (diagonal) and two-parameter joint (off-diagonal) distributions of the posterior using the PCSK emulator (LHD + HPP) in the MCMC for the parameters related to the initial-state rapidity loss. The upper right part (magenta) shows the parameter correlation without the monotonic constraint. The lower left part (green) represents the results imposing the constraint $y_{\rm loss,2}\leq y_{{\rm loss},4}\leq y_{{\rm loss},6}$.}
    \label{fig:yloss_corner}
\end{figure}

The Bayesian analysis usually reports the parameter set $\bs\theta^\mathrm{MAP}$ at the MAP (maximum a posteriori probability) estimate using the GP emulator. In our case, we find the MAP parameter set in the entire prior space gives a reverse ordering for some rapidity loss parameters, namely $y_{\rm loss,4}>y_{\rm loss,6}$. In the Bayesian analysis above, we did not impose any constraints on the prior distribution of the rapidity loss parameters. As shown in Fig.~\ref{fig:posterior_parametrizations}, the experimental measurements at 200 GeV give strong constraints on the initial-state rapidity loss near its beam rapidity, $y_{\rm beam}\equiv\mathrm{arccosh}{\left(\sqrt{s_{\rm NN}}/(2m_{\rm N})\right)}=5.36$. Therefore, the values of $y_{\rm loss,4}$ and $y_{\rm loss,6}$ are anti-correlated (see Fig.~\ref{fig:yloss_corner} below), and both orderings are allowed in the posterior distribution.
Although the MAP parameter set with $y_{\rm loss,4}>y_{\rm loss,6}$ is sufficient to perform simulations for the RHIC BES program, we expect it to fail when extrapolating to the higher LHC energies. Therefore, we choose to report the highest likelihood parameter set with the monotonic constraint $y_{\rm loss,2}\leq y_{\rm loss,4}\leq y_{\rm loss,6}$ in Table~\ref{tab:high_L}. We compute the Bayesian evidence without and with the monotonic constraint and obtain a Bayes factor of $\ln (\mathcal{B}_\mathrm{w.o.\,constr./w.\,constr.}) = 0.85 \pm 0.02$, which falls in the ``inconclusive'' category based on the Jeffreys' Scale~\cite{doi:10.1080/00107510802066753} to distinguish the two setups.

Figure~\ref{fig:posterior_yloss_constraint} compares the posterior distribution for the averaged rapidity loss $\langle y_{\rm loss}\rangle$ as a function of $y_{\rm init}$ with and without the monotonic constraints, $y_{\rm loss,2}\leq y_{{\rm loss},4}\leq y_{{\rm loss},6}$. The 90\% posterior band shrinks almost by half for $y_{\rm init} \in [3, 6]$. Because the values for $y_{\rm loss,2}$ and $y_{{\rm loss},4}$ are also anti-correlated in the posterior distribution, the $\langle y_{\rm loss}\rangle$ at small $y_{\rm init}$ also receive some minor effects from the monotonic constraint.

Figure~\ref{fig:yloss_corner} shows the marginalized posterior distributions for individual rapidity loss parameters on the diagonal and the correlations between the different parameters in the off-diagonal corners.

The diagonal panels show that imposing the monotonic constraint on the rapidity loss parameters leads to more peaked marginalized posterior distributions because the prior space is smaller with the constraint.
While the peak position of $y_{{\rm loss},2}$ is barely affected by the constraint, the peak position of $y_{{\rm loss},4}$ is shifted to a slightly smaller value and the position of $y_{{\rm loss},6}$ to a slightly larger value.
The marginal distribution for the rapidity loss fluctuations $\sigma_{y_{\rm loss}}$ is almost unaffected. We have further checked that the marginalized posterior distributions of the other 16 parameters are not affected by imposing the monotonic constraint.

The upper right off-diagonal panels show the correlation between the rapidity loss parameters without the monotonic constraint.
We find strong anti-correlations between $y_{{\rm loss}, 2}$ and $y_{{\rm loss}, 4}$ and between $y_{{\rm loss}, 4}$ and $y_{{\rm loss}, 6}$.
Such anti-correlation is a consequence of the piece-wise parameterization in Eq.~\eqref{eq:ylossParam} when the experimental measurements at 200 and 19.6 GeV set constraints on the average rapidity loss near their beam rapidity 5.36 and 3.04, respectively.
Looking at the constrained case in the lower left part of the off-diagonal panels, we also observe a sharp cut in the $y_{{\rm loss},6}$ vs. $y_{{\rm loss},4}$ distribution, which is a direct consequence of the $y_{{\rm loss},4}\leq y_{{\rm loss},6}$ constraint in the prior.

\section{Sensitivity Analysis}
\label{sec:sensitivity}

The fast model emulator lets us quantitatively study how individual model parameters affect different experimental observables. 

Starting with a global sensitivity analysis, we compute the first-order Sobol' indices~\cite{sobol1990sensitivity} for individual experimental observables over the entire prior region.
Similar analyses were performed for heavy-ion collisions with (2+1)D simulations assuming longitudinal boost invariance ~\cite{Liyanage:2022byj, Heffernan:2023utr}.
The first-order Sobol' indices quantify the importance of each parameter for a given observable $\mathbb{O}(\bs\theta)$ by performing a decomposition of the variance of $\mathbb{O}(\cdot)$ over the parameter space $\bs\theta \equiv (\theta_1, \dots \theta_m)$.
The first-order Sobol' index for the model parameter $\theta_i$ is defined as
\begin{align}
    S_i^\mathbb{O} \equiv \frac{\mathrm{Var}_{\theta_i}\left(\mathbb{E}_{\bs\theta_{-i}}(\mathbb{O}(\bs\theta)\vert \theta_i)\right)}{\mathrm{Var}_{\bs\theta}(\mathbb{O}(\bs\theta))},\quad i=1,\dots,m.
\end{align}
Here, the index $i$ runs over the dimension of the model parameter space $(m = 20)$. We sample each model parameter $\theta_i$ independently from a uniform distribution over its corresponding prior range in Table~\ref{tab:parameters}. 
The expression $\mathbb{E}_{\bs\theta_{-i}}(\mathbb{O}(\bs\theta)\vert \theta_i)$ performs an average of the observable $\mathbb{O}$ over a parameter subspace $\bs\theta_{-i} \equiv \bs\theta \backslash \theta_i$ by holding the $i$-th parameter $\theta_i$ at a fixed value.
Then Sobol' index $S_i^\mathbb{O}$ measures the fraction of the variance of the observable $\mathbb{O}$ from the parameter $\theta_i$ relative to its total variance.

\begin{figure*}[t!]
    \centering
    \includegraphics[width=\textwidth]{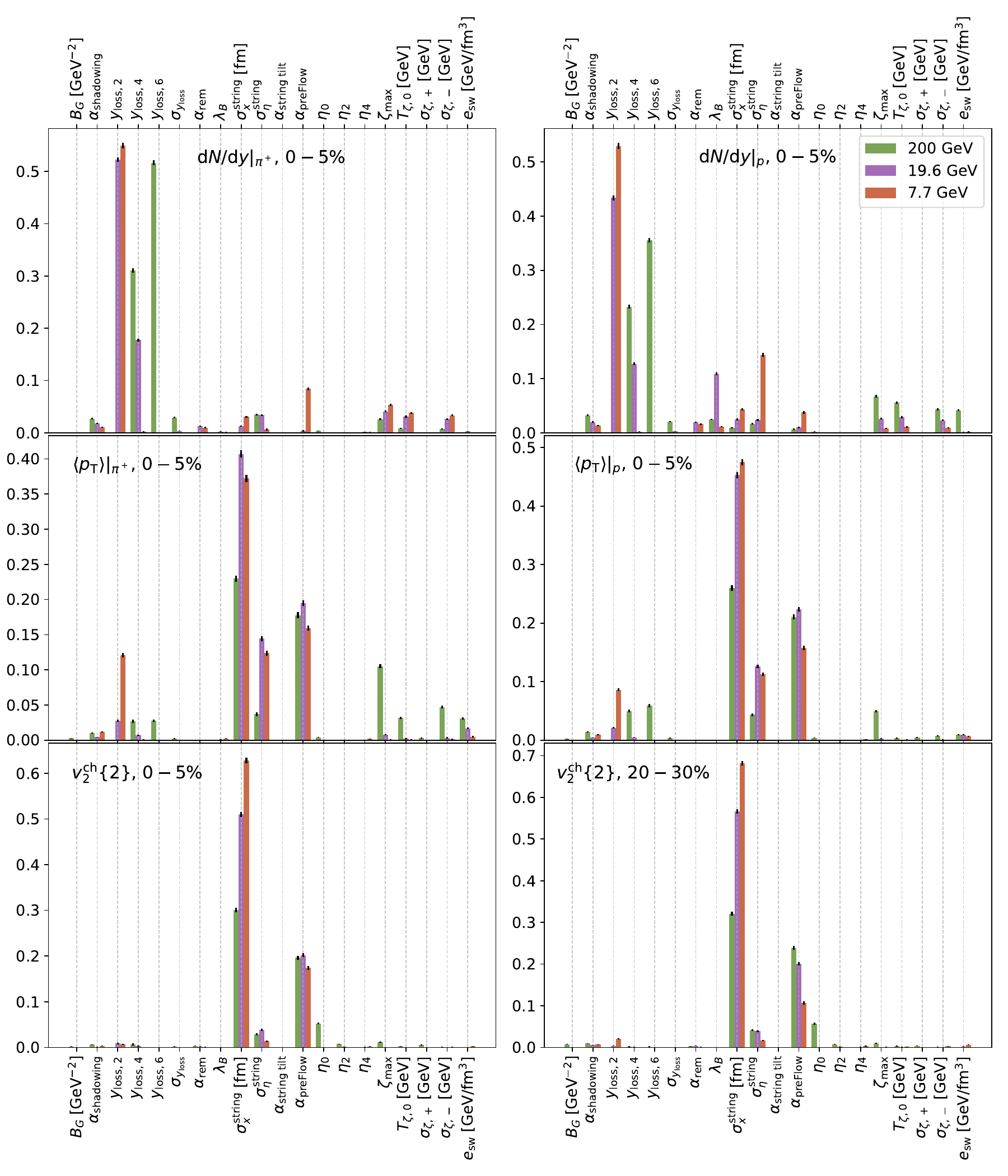}
    \caption{The first-order Sobol' indices for selected midrapidity observables at different collision energies $\sqrt{s_{\rm NN}}$ and centrality bins. The black bars indicate the 95\% confidence interval.}
    \label{fig:sensitivity_Sobol1}
\end{figure*}

Figure~\ref{fig:sensitivity_Sobol1} shows the first-order Sobol' indices for identified particle yields $\mathrm{d}N/\mathrm{d}y$ and $\langle p_{\rm T}\rangle$ in the most central events and charged hadron $v_{2}^{\rm ch}\{2\}$ at 0-5\% and 20-30\% centrality bins.
The particle yields of charged pions have strong sensitivities to the rapidity loss parameters $y_{\mathrm{loss}, n}$. The measurements at 200 GeV are sensitive to the rapidity loss parameters at incoming rapidity $y_\mathrm{init} = 4$ and $y_\mathrm{init} = 6$, while the strongest sensitivity shifts to $y_\mathrm{loss, 2}$ for the lower collision energy measurements. Comparing the Sobol' indices between pion and proton yields, we find that the proton yields show additional sensitivity to the baryon stopping parameter $\lambda_B$ and the string's longitudinal profile parameter $\sigma^\mathrm{string}_\eta$.

The averaged transverse momenta of pions and protons are mostly sensitive to the initial-state hotspot size in the transverse plane $\sigma^\mathrm{string}_x$ and the magnitude of pre-equilibrium flow $\alpha_\mathrm{preFlow}$.
Both model parameters have strong effects on the development of hydrodynamic radial flow. 
The hotspot's transverse size controls the magnitude of the early-stage pressure gradients.
A stronger pre-equilibrium flow leads to faster fireball expansion at the early time. 
We also find the pion's mean $p_{\rm T}$ at 200 GeV shows some sensitivity to the specific bulk viscosity of the QGP. 
However, the magnitudes of their Sobol' indices are dwarfed compared to those model parameters in the initial-state and pre-equilibrium stages.
A similar situation is observed for mid-rapidity anisotropic flow. 
The charged hadron elliptic flow coefficients at central and semi-peripheral collisions are mostly sensitive to the hotspot's transverse size and pre-equilibrium flow over the entire prior range. 
The sensitivity to the specific shear viscosity of the QGP is very small. 
We checked that the charged hadron triangular flow shows very similar global sensitivity as the elliptic flow observables.

\begin{figure}[h!]
    \centering
    \includegraphics[width=\linewidth]{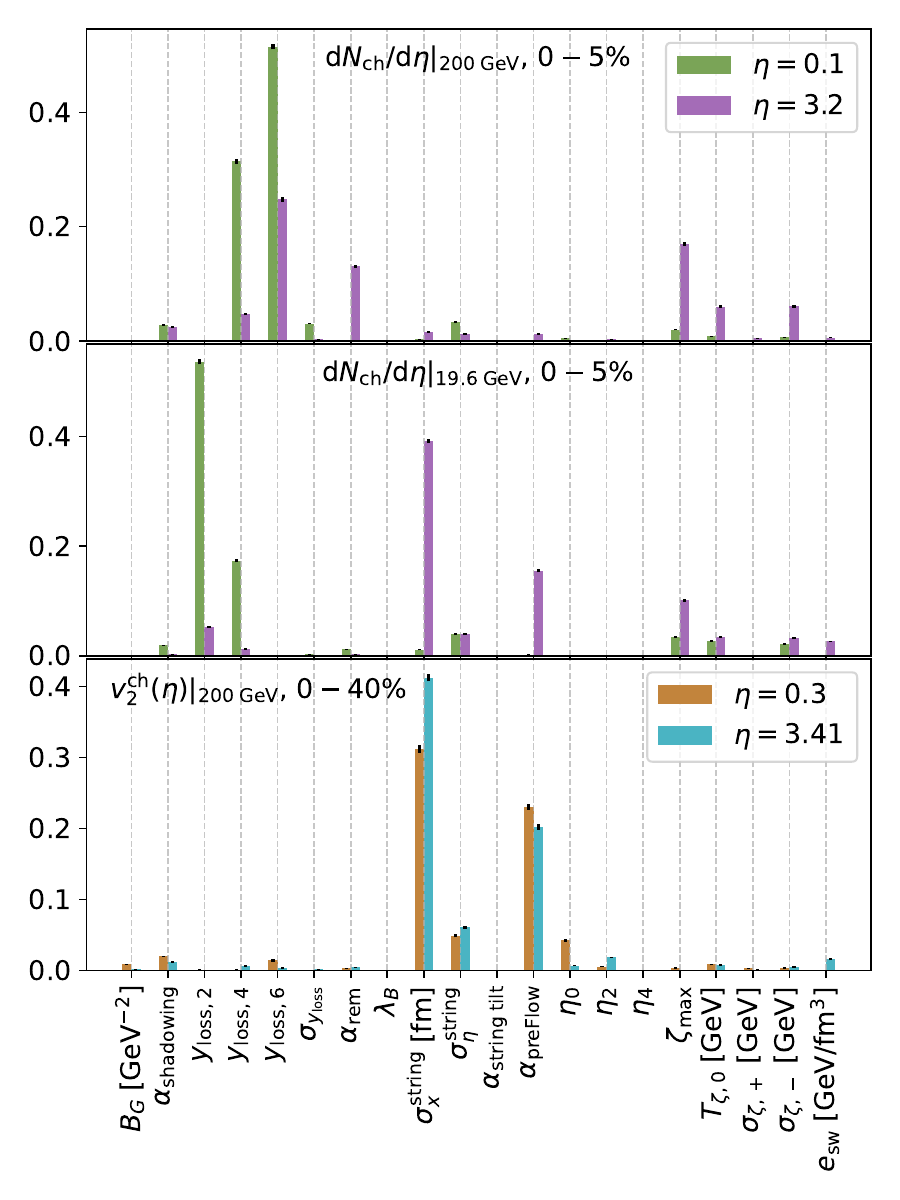}
    \caption{The first-order Sobol' indices for selected observables at central and forward rapidity regions. The black bars indicate the 95\% confidence interval.}
    \label{fig:sensitivity_Sobol2}
\end{figure}

Figure~\ref{fig:sensitivity_Sobol2} further represents the first-order Sobol' indices for rapidity-dependent observables. 
We find that the charged hadron yields at forward rapidity show a stronger sensitivity to the specific bulk viscosity of the QGP and also the amount of rapidity loss in the beam remnant $\alpha_\mathrm{rem}$.
We find a similar global sensitivity for charged hadron elliptic flow at central and forward rapidity regions.

Although it is informative to quantify the global sensitivity of experimental observables to the model parameter over the entire prior parameter space shown in Figs.~\ref{fig:sensitivity_Sobol1} and~\ref{fig:sensitivity_Sobol2}, the responses could be different in different local regions of the parameter space. Therefore, we further compute the response coefficients between experimental observables and model parameters using the posterior distribution. We follow the approach in Ref.~\cite{Sangaline:2015isa} and define the ensemble-averaged response coefficient matrix,
\begin{align}
    \mathcal{R}_{ai} \equiv \left\langle \frac{\partial y_a}{\partial \theta_i} \right\rangle_\mathrm{post} = \sum_j \langle \delta y_a \delta \theta_j \rangle_\mathrm{post} \langle \delta \theta_j \delta \theta_i \rangle_\mathrm{post}^{-1}.
    \label{eq:responseMatrix}
\end{align}
Here, the index $a$ runs over individual experimental observables in Table~\ref{tab:training_data}, and the index $i$ runs over the model parameters in Table~\ref{tab:parameters}. The ensemble average $\langle \cdot \rangle_\mathrm{post}$ is performed on the posterior distribution to extract the model's response to the experimental observables near the real measurements. To make the response coefficients unitless, we define the normalized experimental observables $\delta y_a = (y_a(\theta_j) - \bar{y}_a)/\sigma_{y_a}$ and model parameters $\delta \theta_i = (\theta_i - \bar{\theta}_i)/\sigma_{\theta_i}$, where $\bar{y}_a$($\bar{\theta}_i$) and $\sigma_{y_a}$($\sigma_{\theta_i}$) are the mean and standard deviation of the observable $y_a$ (parameter $\theta_i$) over the posterior distribution, respectively. The right-hand side of Eq.~\eqref{eq:responseMatrix} should be interpreted as a matrix multiplication. 

\begin{figure*}[t!]
    \centering
    \includegraphics[width=\textwidth]{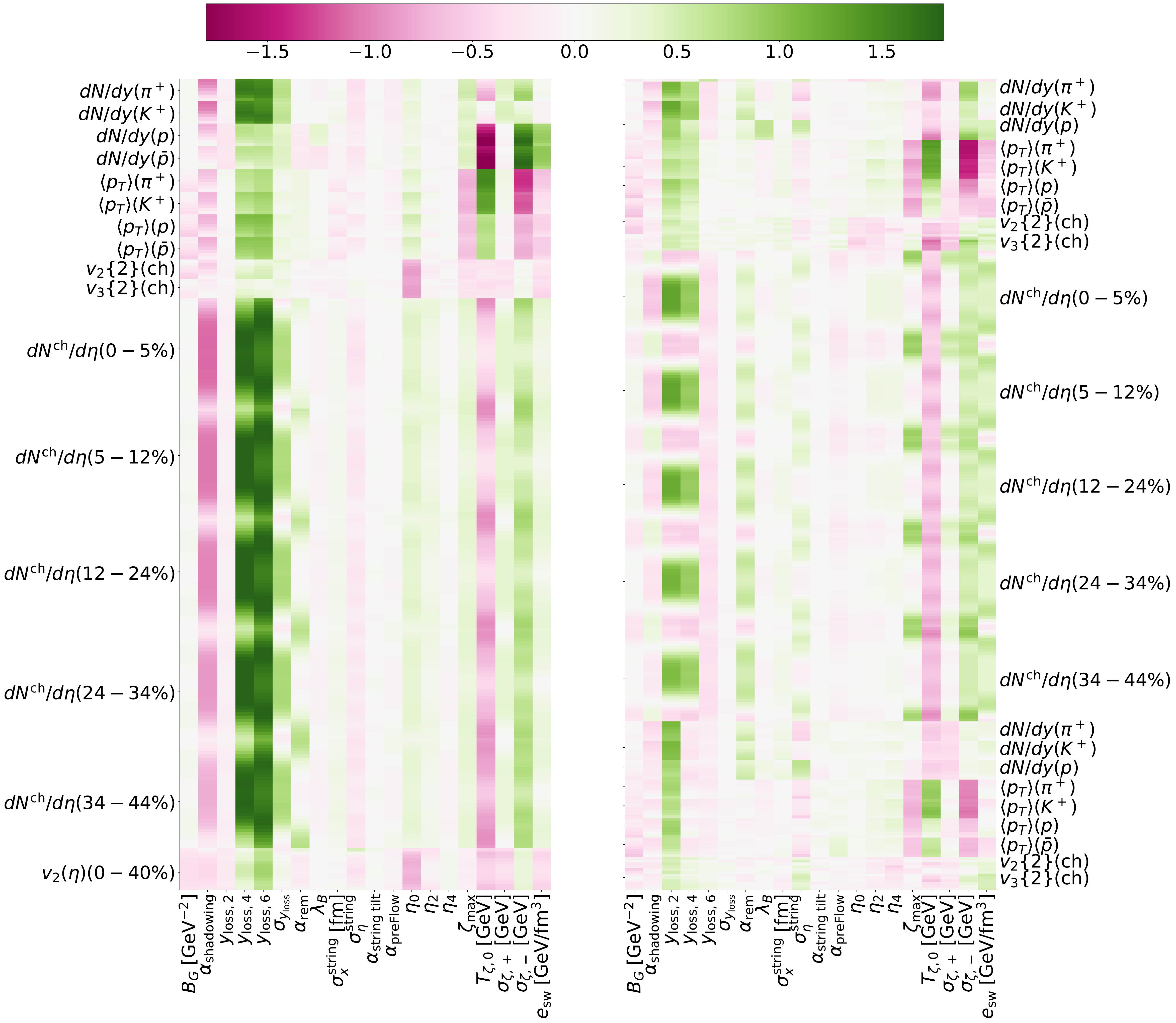}
    \caption{Response matrix $\mathcal{R}_{ai}$ averaged over the posterior distribution. The observables in the left column correspond to observables at $\sqrt{s_{\rm NN}}=200\;\mathrm{GeV}$ collisions. The right column shows the response to observables at $\sqrt{s_{\rm NN}}=19.6\;\mathrm{GeV}$ and $7.7\;\mathrm{GeV}$.}
    \label{fig:sensitivityPost}
\end{figure*}

Figure~\ref{fig:sensitivityPost} shows the response matrix between the experimental observables and model parameters over the posterior distribution. We observe that identified particle yields show strong sensitivity to the rapidity loss parameters, consistent with the results from the global analysis above. The signs of the response coefficients provide additional information compared to Sobol' indices. Interestingly, the mid-rapidity charged hadron yields at 19.6 GeV show positive correlations with the rapidity loss parameters while the forward ${\rm d}N^\mathrm{ch}/{\rm d}\eta$ negatively correlates with these parameters. This pattern results from the energy-momentum conservation imposed in the \texttt{3D-Glauber} model. More particle production near mid-rapidity would require more energy loss during the initial state, which leaves less available energy at forward rapidity regions. Particle yields show significant anti-correlation with the parameter $\alpha_\mathrm{shadowing}$ because a smaller value of $\alpha_\mathrm{shadowing}$ produces more string sources from the binary collisions and increases mid-rapidity particle production.

Different from the global sensitivity analysis above, we find the observables' response coefficients to the hotspot transverse size $\sigma^\mathrm{string}_x$ and pre-equilibrium flow $\alpha_\mathrm{preFlow}$ are small in the posterior distribution. This difference is because the marginalized posterior distributions for these two parameters are sharply peaked around the edge of their prior range, as shown in Fig.~\ref{fig:posterior_comparison}. The correlations in the local region are small and different from the global averages. 

We find that the observables' responses to the QGP transport properties in Fig.~\ref{fig:sensitivityPost} are more in line with our physics intuition. 

The identified particles' mean $p_{\rm T}$ shows sizable negative correlations with the parameters related to the specific bulk viscosity of the QGP. This result aligns with our intuition that the bulk viscosity acts as a resistance to reduce the development of hydrodynamic radial flow.

The parameters related to the specific shear viscosity of the QGP have a significant negative correlation with the charged hadrons' anisotropic flow coefficients at 200 GeV. The pseudo-rapidity dependent $v_2(\eta)$ measurements also show strong sensitivity to the QGP shear viscosity, especially the $v_2$ at forward rapidity shows some sensitivity to the $\eta_2$ parameter. Although the correlation is relatively weak, we find a negative correlation between charged hadron $v_n$ at 19.6 GeV and the shear viscosity at $\mu_B = 200$~MeV. And the $\eta_4$ parameter, which controls the QGP shear viscosity at $\mu_B = 400$~MeV, is negatively correlated with the charged hadron $v_n$ at 7.7 GeV.

\section{Conclusion}
\label{sec:conclusion}
This work presents a systematic Bayesian analysis of particle production and flow measurements at the RHIC BES program using an event-by-event (3+1)D hydrodynamics + hadronic transport framework. We obtain robust statistical constraints on the QGP transport properties and various aspects of relativistic nuclear dynamics.
This work expands upon Ref.~\cite{Shen:2023awv} and provides the complete 20-dimensional model posterior distribution using an improved Gaussian Process model emulator. Utilizing the statistical tools developed in Ref.~\cite{Roch:2024xhh}, we systematically analyze the model's posterior distributions obtained from three model emulators with different accuracy. Using the statistical measures, namely the Kullback-Leibler divergence and Bayes factor, we demonstrate that the accuracy of the model emulator plays a crucial role in getting robust posterior distribution from the experimental measurements.
With an independent Bayesian analysis framework~\cite{hendrik_roch_2024_12807892}, we verify that the posterior distributions of the initial-state rapidity loss and the specific shear and bulk viscosity of QGP obtained in this work are consistent with the results in Ref.~\cite{Shen:2023awv}, demonstrating the reproducibility of our results.

We sample 100 parameter sets from the posterior distributions and perform large-scale event-by-event simulations at each parameter set using our (3+1)D hybrid model. These results verify that the posterior distribution obtained with the model emulator is reliable for the real model. We further make predictions for a set of $p_{\rm T}$-differential observables and estimate their systematical errors based on the variations among the 100 parameter sets. Good agreements with the experimental data are found. Based on this approach, we will perform more systematic model predictions in future work.
For community users who do not have sufficient computing resources to perform simulations at multiple posterior parameter sets, we provide a high-likelihood parameter set in Table~\ref{tab:high_L}.

Lastly, we report a sensitivity analysis of how individual experimental observables respond to different model parameters, enabled by the fast model emulator. We find commons and differences between the global sensitivity analysis over the entire prior region and the response matrix from the posterior distribution. Both sensitivity analyses provide useful physics insights into the phenomenological modeling of relativistic heavy-ion collisions. We will extend the sensitivity analysis to a more systematic experimental design study in future work.

We provide the following results from this work for community use:
\begin{itemize}
    \item Training datasets in Table~\ref{tab:training_data}~\cite{jahan_2024_12807556}
    \item Trained model emulator PCSK (LHD + HPP)~\cite{jahan_2024_12807556}
    \item Posterior parameter samples from the MCMC~\cite{jahan_2024_12807556}
    \item An updated Bayesian analysis package with the \texttt{pocoMC} integration~\cite{hendrik_roch_2024_12807892}
\end{itemize}

\begin{acknowledgments}
We thank Bj\"{o}rn Schenke and Wenbin Zhao for their involvement during the early stage of the project and useful discussions.
This work is supported in part by the U.S. Department of Energy, Office of Science, Office of Nuclear Physics, under DOE Award No.~DE-SC0021969 and DE-SC0024232. H.~R. and C.~S. were supported in part by the National Science Foundation (NSF) within the framework of the JETSCAPE collaboration (OAC-2004571).
C.~S. acknowledges a DOE Office of Science Early Career Award.
S.-A.~J. acknowledges support from the BAND Fellowship funded by the National Science Foundation CSSI program under award number OAC-2004601. Numerical simulations presented in this work were partly performed at the Wayne State Grid, and we gratefully acknowledge their support.
This research was done using resources provided by the Open Science Grid (OSG)~\cite{Pordes:2007zzb, Sfiligoi:2009cct, https://doi.org/10.21231/906p-4d78, https://doi.org/10.21231/0kvz-ve57}, which is supported by the National Science Foundation award \#2030508 and \#1836650.
\end{acknowledgments}

\bibliography{bib, non-inspire}

\end{document}